\documentclass[twocolumn,superscriptaddress,prb]{revtex4}
\usepackage{amssymb}
\usepackage{graphicx}
\usepackage{dcolumn}
\usepackage{bm}

\begin{document}
\title{The highly tunable Mg-Ni-H switchable mirror system}
\author{J.L.M. van Mechelen}\email[Present address: D\'epartement de Physique de la Mati\`ere Condens\'ee, University
of Geneva, Quai Ernest-Ansermet 24, 1211 Gen\`eve, Switzerland.\\ Email : ] {Dook.vanMechelen@physics.unige.ch}
\affiliation{Faculty of Sciences, Department of Physics and Astronomy, Condensed Matter Physics, Vrije
Universiteit, De Boelelaan 1081, 1081 HV Amsterdam, The Netherlands}

\begin{abstract}
The Mg-Ni-H switchable mirror system shows a spectacular composition dependence in the hydrogenated state. The
optical transmission of a hydrogenated Mg-Ni-H gradient film displays a stripe-like pattern that consists of
alternating regions of high and low transmission. This behavior is peculiar as the composite Mg-Ni-H phase
diagram only predicts a gradual change between MgH$_2$ and Mg$_2$NiH$_4$ which can not account for the observed
composition dependence. In order to understand the observed transmission pattern, the Mg-Ni-H system has been
investigated optically, electrically, structurally and morphologically in the metallic and hydrogenated state.
As-deposited, metallic, Mg-Ni composition gradient thin films contain the Mg$_2$Ni and MgNi$_2$ phases for
well-defined composition regions. In the hydrogenated state, vibrational spectroscopy shows the composition
dependence of semiconducting Mg$_2$NiH$_4$ which is formed from metallic Mg$_2$Ni. On exposure to hydrogen,
physical properties change as well near the equiatomic composition and the composition of MgNi$_2$, pointing to
possible hydride formation. The strong composition dependence of the Mg-Ni-H system, expressed in the observed
transmission, can eventually be understood from the distribution of Mg$_2$NiH$_4$ and the occurrence of MgH$_2$
for higher Mg contents.
\end{abstract}

\maketitle

\section{Introduction} In 1996 Huiberts \textit{et al.}~\cite{Huiberts} discovered the dazzling optical changes of
yttrium and lanthanum upon the absorption of hydrogen. On exposure to hydrogen, a switchable mirror thin film
changes from reflecting to transparent while undergoing a metal-to-insulator transition. All rare earth and Mg
alloyed rare earth metal hydrides~\cite{Sluis} possess these properties. In 2001, Richardson \textit{et
al.}~\cite{Richardson} found that alloys composed of Mg and transition metals like Ni, Co, Fe and Mn also switch
reversibly. Among these rare earth free metal hydrides, Mg$_2$NiH$_4$ received special attention because of its
hydrogen storage capacity with an improved kinetics compared to MgH$_2$ and because of its possible application
as smart coating~\cite{VanMechelen}.

In the metallic Mg-Ni system, the Mg-Ni bulk phase diagram shows that Mg$_{2}$Ni and MgNi$_{2}$ are the two
existing intermetallic phases at room temperature.~\cite{PhaseDiagram} Mg$_{2}$Ni can absorb hydrogen, whereas it
has not been reported for MgNi$_{2}$. During hydrogenation, the metal solid solution Mg$_{2}$NiH$_{0.3}$ is first
formed. On increasing the hydrogen content, the host metal atoms rearrange and a complex hydride structure
Mg$_{2}$NiH$_{4}$ is created that consists of [NiH$_4$]$^{4-}$ complexes that are ionically bound to Mg$^{2+}$.
Compared to Mg$_{2}$Ni, the stoichiometric Mg$_2$NiH$_4$ structure has been expanded by 32 vol.\%.~\cite{Schefer}
For $T>510$ K, Mg$_2$NiH$_4$ has a cubic structure which becomes monoclinally distorted below this temperature.

\begin{figure}
\includegraphics[width=\linewidth]{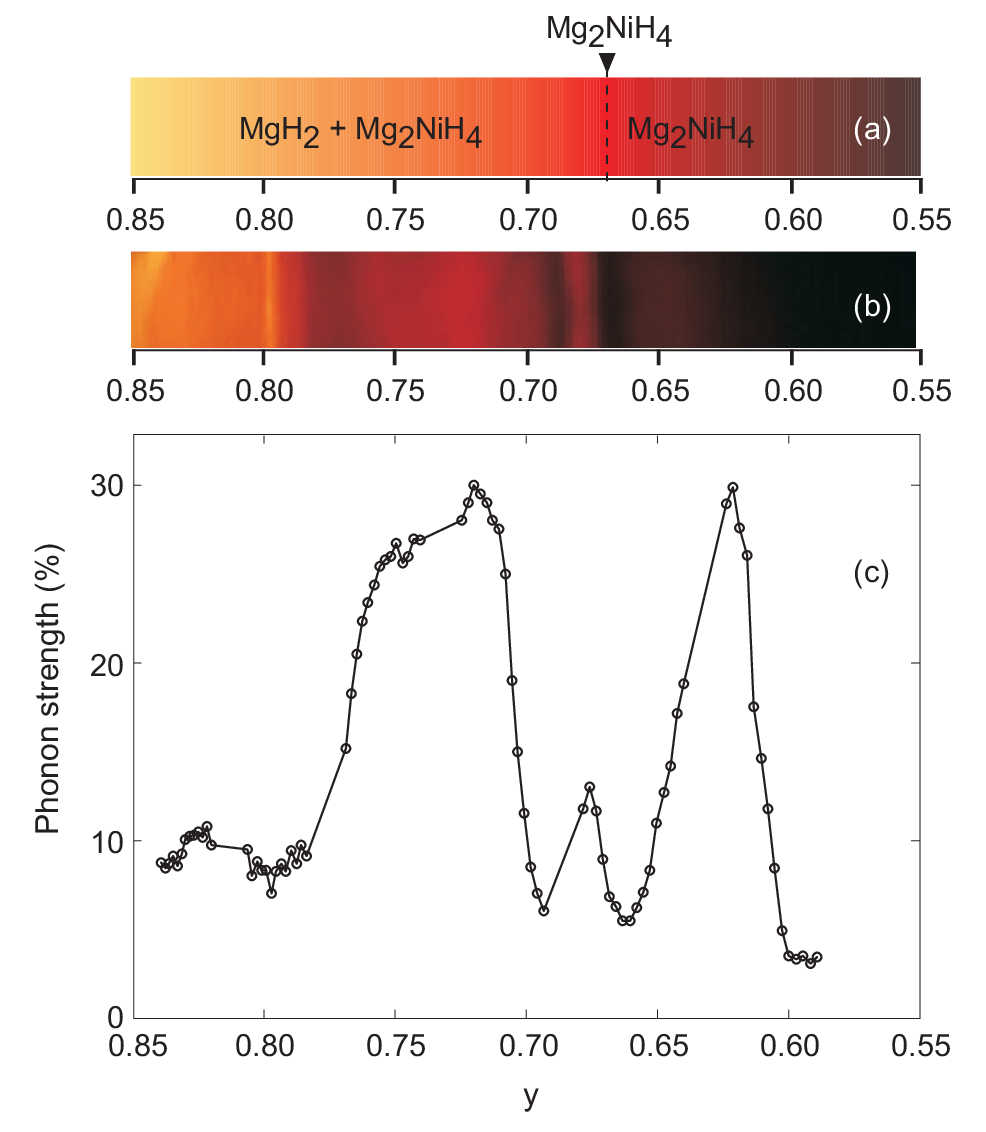}
\caption{\label{IntroductionPicture} Illustration showing the inconsistency between (a) the predicted
(linearized) Mg-Ni-H phase diagram as a function of composition $y$ of Mg$_y$Ni$_{1-y}$H$_x$ and (b) the camera
recorded visible transmission of a hydrogenated Mg-Ni-H gradient thin film, and on the other hand the excellent
correlation of the transmission (b) with (c) the phonon strength of the [NiH$_4]^{4-}$ cluster, indicating the
amount of formed Mg$_2$NiH$_4$ and therefore the level of optical transmission due to Mg$_2$NiH$_4$.}
\end{figure}

For the hydrogenated Mg-Ni-H system, the ternary bulk Mg-Ni-H phase diagram has been calculated thermodynamically
from experimental data of the binary constituent systems Mg-H, Ni-H, and Mg-Ni.~\cite{Zeng} The predicted phase
diagram shows that the only hydrides that can be formed are Mg$_2$NiH$_4$ and MgH$_2$.
Fig.~\ref{IntroductionPicture}a displays a linearized representation of this phase diagram, where metal
inclusions like Mg$_2$Ni, MgNi$_2$, Mg$_2$NiH$_{0.3}$ and Ni are present next to the two hydrides. This
corresponds to a non-complete hydrogen loading as observed in practice~\cite{Enache}. At Mg-rich compositions,
the physical properties are dominated by MgH$_2$ which for decreasing Mg contents is \textit{gradually} replaced
by Mg$_2$NiH$_4$. The depicted colors are based on the fact that Mg$_{0.9}$Ni$_{0.1}$H$_x$ has a yellow visual
appearance,~\cite{Richardson} Mg$_2$NiH$_4$ is red~\cite{Reilly} and MgNi$_2$ does not react with hydrogen and
thus remains a non-transparent metal which is imaged as black.

In order to study composition dependent systems, it is desirable to monitor a continuous range of compositions at
the same time. This can be achieved by making thin film samples with a composition gradient for which the element
ratio is a monotonic function of the position on the sample. In this way, Mg-Ni composition gradient samples are
deposited from the elements. Fig.~\ref{IntroductionPicture}b shows the optical transmission of a Mg-Ni-H
composition gradient sample capped with 5 nm Pd on a glass substrate. The sample is hydrogenated in an optical
gas loading cell at 50 $^\circ$C and p$_{H_2}=2\cdot10^5$ Pa. The composition $y$ of Mg$_y$Ni$_{1-y}$ changes
from left to right from 0.85 to 0.55, respectively. When illuminated with white light, a strongly composition
dependent stripe-like pattern perpendicular to the gradient becomes visible in transmission. Regions of high
transmission alternate with zones of low transmission, while the color mainly changes from brilliant yellow at
the Mg rich part to reddish black at the Mg poor side. On the basis of the predicted Mg-Ni-H phase diagram, which
indicates only a gradual change between MgH$_2$ and Mg$_2$NiH$_4$ in this composition range, it is impossible to
explain the observed composition dependence.

Although the predicted Mg-Ni-H phase diagram can not account for the observed transmission behavior, several
physical properties correlate very well. An example is the optical phonon strength of the [NiH$_4]^{4-}$ cluster
of Mg$_2$NiH$_4$ displayed as a function of composition (Fig.~\ref{IntroductionPicture}c). The phonon strength is
a direct measure for the amount of formed Mg$_2$NiH$_4$ and so for the level of transmission due to
Mg$_2$NiH$_4$. Comparison with the optical transmission shows the excellent resemblance between both properties
for compositions $y\lesssim 0.8$.

In summary, the intriguing Mg-Ni-H system does not behave following the predicted Mg-Ni-H phase diagram, but many
of its physical properties correlate in a way possible to draw a picture about it. In order to understand the
hydrogenated system, first the as-deposited Mg-Ni system is optically, electrically, structurally and
morphologically investigated, as reported in section~\ref{MgNi}. Similarities and correlations between the many
metallic properties clearly show the presence of site-ordered and structurally disordered intermetallic phases in
well-defined composition ranges. Then, section~\ref{MgNiH} reports about the hydrogenated Mg-Ni-H system. The
observed stripe-like pattern in transmission is quantified by photometric spectrometry in order to compare it to
other optical, electrical and structural phenomena. Correlations between these properties show the stripe-like
pattern to be due to a peculiar spatial distribution of MgH$_2$ and Mg$_2$NiH$_4$ together with an enhanced
presence of Mg$_2$NiH$_{0.3}$ at certain compositions. Eventually, the influence of hydrogen on the intermetallic
phases is demonstrated by the DC resistivity in the framework of an effective medium theory (Bruggeman
approximation) and classical percolation theory.


\section{Experimental}
\subsection{Film deposition} Mg-Ni films with a compositional gradient are deposited by magnetron co-sputtering in
a high-vacuum chamber (base pressure $\sim10^{-6}$ Pa) at room temperature. The mutual power of the Mg and Ni
particle current and their direction with respect to the substrate control the composition regime of the film. As
a consequence of the technique, films possess a thickness gradient, which is measured \textit{ex situ} with a
Dektak stylus profilometer. Typically, the gradient is almost linear and varies from 300 to 600 nm, increasing
towards the Mg rich side. To catalyze the hydrogen uptake and to protect the film from oxidation a homogeneous 4
nm thin Pd cap layer is deposited \textit{in situ} on top of the Mg-Ni film. Mg-Ni gradient films are either
deposited on a 70 mm long substrate made of one piece or on seven 10 mm long substrates that are placed next to
each other.

\subsection{Composition determination} The composition of as-deposited gradient samples is determined \textit{ex
situ} by Wavelength Dispersive X-ray Spectrometry (WDS) and Rutherford Backscattering Spectrometry (RBS). WDS has
been carried out in a JEOL JXA-8800M Electron Microprobe. Here characteristic X-rays are produced by the
interaction of a 10 $\mu$m electron beam with the film layer. In order to have the excitation volume mainly
within the thin film layer, the acceleration voltage of the beam is adjusted to the thickness and the density of
the layer. In this way, the microprobe was operated at 5 kV and 15 kV which allowed the detection of the
K$_\alpha$ line of Mg and the L$_{\alpha}$ and K$_{\alpha}$ line of Ni, respectively. In order to diminish the
background radiation glassy carbon substrates are used, which low energy X-rays can not detect. Most samples for
composition analysis did not have the highly absorbing Pd cap layer.

For RBS, $^4$He$^+$ ions with an energy of 2 MeV bombard the sample and the energy distribution of the
backscattered particles within a solid angle d$\Omega$ is detected. The maximum energy of the scattered ions
determines the mass of the scattering centers, from which the present elements and their amounts can be
established. In order to diminish the background radiation, also for RBS glassy carbon substrates are used.

WDS makes use of predefined calibration standards and has to apply a matrix correction after the measurement
whereas RBS directly measures the backscattered radiation of atomic fractions. On the other hand, the electron
beam size of WDS enables the detection of composition irregularities on a very small scale, while for RBS the
diameter of the $\alpha$ particle beam is about $1-2$ mm. Besides the Mg and Ni amount, RBS analysis also
provides the oxygen content of the sample.

For Mg$_{y}$Ni$_{1-y}$ compositions with $y<0.83$ WDS and RBS agree within 5\%, but the deviation increases at
higher Mg contents. Further, in the direction of the composition gradient, WDS showed composition deviations with
respect to the gradual composition change on a 250 $\mu$m scale, to vary from 0.0067 at $y=0.86$ to 0.0029 at
$y=0.29$. This indicates a very smooth composition change which is mainly determined by the composition gradient
on purpose.

\subsection{Optical measurements} Optical reflection and transmission measurements are performed in two
overlapping energy ranges using two Bruker IFS66 Fourier transform infrared spectrometers. The first range
extends from 0.72 to 3.5 eV (i.e., $1722-354$ nm) and covers the visible and near infrared, the second range is
between 62 meV and 1.2 eV (i.e., $20-1.0\ \mu$m) which covers the near, middle and beginning of the far infrared.
Hydrogen loading is performed \textit{in situ} in an optical gas loading cell. As the cell could only accommodate
10 mm long samples, seven similar hydrogen loadings were required to obtain information about one long gradient
sample. During loading the temperature is below 50 $^\circ$C and the hydrogen pressure is increased from $10^2$
Pa to $2\times10^5$ Pa. Simultaneously, the mean electrical resistivity over the present composition gradient has
been monitored. As the visible and infrared beam diameter at the sample is about 3 mm, optical properties are
averaged over only a narrow composition range. Therefore, the reflection in the as-deposited state and the
reflection and transmission in the final state could be measured through making line scans perpendicular to the
composition gradient.

\subsection{Electrical measurements} The electrical resistivity has been measured in a home-made apparatus which
enabled resistivity measurements on gradient samples. The apparatus has a transparent Perspex arm on which four
gold, spring damped needles with a tip diameter of 25 $\mu$m are mounted in a $2\times2$ mm$^2$ square. On one
side, an axis pierces into the arm along which the arm can slide perpendicular to the composition gradient of the
sample and around which it can rotate to place the needles on the sample. The gradient sample has zones of 2.5 mm
that are electrically isolated by stripes of 0.5 mm which are shadowed during deposition. The empty stripes are
necessary to confine the current to the desired composition range. In this way the electrical resistivity has
been measured by the Van der Pauw method~\cite{Pauw} which eliminates the additional resistance coming from
objects on the electrical path different than the film.

Resistivity measurements of as-deposited gradient samples are carried out \textit{ex situ}. As resistivity
changes enormously during the first part of the hydrogen desorption, measurements on hydrided gradient samples
are performed in a glove box which contained a mixture of nitrogen and hydrogen.

\subsection{Structural measurements} X-ray diffraction measurements on as-deposited and hydrogenated gradient
samples are performed in $\theta-2\theta$ geometry using a Bruker D8 Discover X-ray diffractometer. Line scans
perpendicular to the composition gradient are carried out with a scan time of 60 s per $\delta2\theta=0.01^\circ$
using a $5-6$ mm Cu K$_{\alpha}$ beam of mean wavelength $\lambda=1.5418$ \AA. A hydrogen filled gas cell,
consisting of a partially X-ray transparent beryllium dome, prevented the pre-hydrogenated sample from unloading.

\subsection{Surface morphology} The surface structure has been determined using a JEOL JSM-6301F Scanning Electron
Microscope (SEM) and a NanoScope III Atomic Force Microscope (AFM). The SEM operated at an acceleration voltage
of 4 kV. The AFM measured in tapping mode using silicon cantilevers. The surface of an as-deposited 70 mm long
gradient sample has first been surveyed at a magnification of 40,000 times by SEM. In order to quantify
morphology parameters like the surface roughness and the grain size, identical composition regions are measured
by AFM.

\section{The M$\text{g}$-N$\text{i}$ binary alloy system}~\label{MgNi} In this section we will consider the
metallic Mg-Ni system, forming the basis of the hydrogenated Mg-Ni-H system. For this purpose as-deposited
composition gradient samples are investigated at room temperature. Optical, electrical and structural properties
indicate the presence of intermetallic phases and their arrangement in the crystal lattice. Structural probes
further describe the interior of the metallic film. Finally, the surface morphology of the metallic layer
illustrates the presence of intermetallic phase formation and complements the structural picture.

\subsection{Optical properties}\label{MetalOptical}
\begin{figure}
\includegraphics[width=\linewidth]{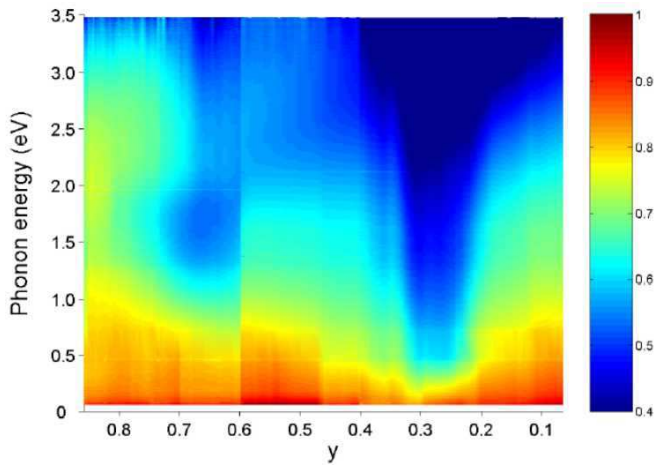}
\caption{\label{MetalReflSurface}Optical reflection of Mg$_y$Ni$_{1-y}$ as a function of energy and composition.
The occurrence of Mg$_2$Ni is characterized around $y=0.68$ by a minimum at $\hbar\omega=1.7$ eV; that of
MgNi$_2$ around $y=0.3$ by reduced values for most visible photon energies. The irregular transition at $y=0.60$
is due to a slightly different alignment of samples for $y>0.60$ and $y<0.60$.}
\end{figure}
\begin{figure}
\includegraphics[width=\linewidth]{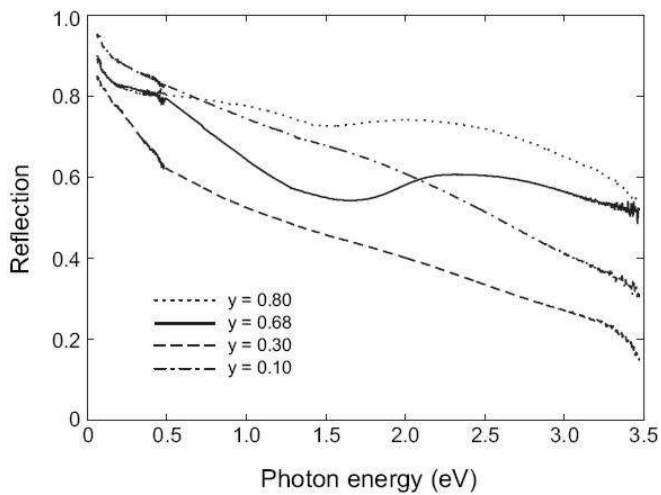}
\caption{\label{MetalReflLines}Optical reflection of Mg$_y$Ni$_{1-y}$ for some selected compositions. Clearly
visible are the interband absorption for $y=0.68$ at $\hbar\omega=1.7$ eV and the significant lowered screened
$\omega_p$ for $y=0.30$.}
\end{figure}

In order to determine the presence of possibly formed phases in as-deposited Mg-Ni composition gradient thin
films, optical reflection measurements were performed for Mg$_y$Ni$_{1-y}$ compositions in the range
$0.07<y<0.86$ on glassy carbon substrates (Figs.~\ref{MetalReflSurface} and \ref{MetalReflLines}).

The reflection increases slightly on the addition of Mg and varies at for instance $\hbar\omega=2$ eV from 75\%
at $y=0.86$ to 65\% at $y=0.07$. This can be understood from the fact that only very weak absorption processes
are present in alkali metals and alkali earths (like Mg), which have only s electrons is the valance band. They
therefore follow the Drude model well, which predicts a high reflection up to the plasma frequency $\omega_p$. On
the other hand, between parallel d bands in Ni interband absorption plays a significant role and the reflectivity
is lowered before $\omega_p$.

The reflection measurements show two regions of much lower intensity around $y=0.68$ and 0.3, which correspond
well to the compositions of the stoichiometric alloys Mg$_2$Ni and MgNi$_2$, respectively. However, as can be
seen from Fig.~\ref{MetalReflSurface}, both composition regions have different spectral features. Around $y=0.33$
the screened plasma frequency is significantly lowered for a very broad composition range, which results in a
strongly reduced reflection down to infrared energies. Most probably this is caused by various interband
absorption processes due to MgNi$_2$. The minimum around $y=0.68$ is different in a sense that it extends over a
much smaller energy range. This dip around $\hbar\omega=1.7$ eV indicates the presence of well-determined
interband absorption at this photon energy. The absorption becomes broader and stronger as $y$ approaches 0.68
and forms a very well defined minimum both in energy and composition (cf. Fig.~\ref{MetalReflSurface}). As will
be demonstrated in the next sections, the local reflection minimum around $y=0.68$ is due to the site-ordered
Mg$_2$Ni phase, whereas the minimum around $y=0.33$ is caused by the structurally disordered MgNi$_2$ phase.

From the optical reflection, the dielectric function $\varepsilon(\omega)=\varepsilon_1+i\varepsilon_2$ can be
parameterized with a Drude-Lorentz model,
\begin{equation}\label{DrudeLorentz}
    \varepsilon(\omega)=\varepsilon_{\infty}-\frac{\omega_p^2}{\omega^2+i\omega/\tau}+\sum_{j=1}^N\frac{f_j}{\omega^2_{0,j}-\omega^2-i\omega\beta_j}.
\end{equation}
\begin{figure}
\includegraphics[width=\linewidth]{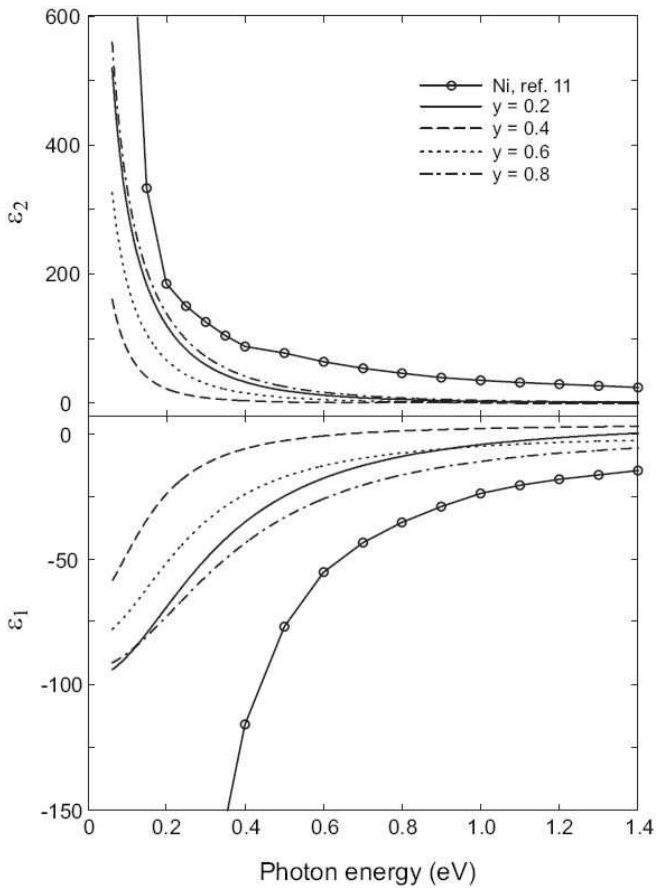}
\caption{\label{MetalEps} Real ($\varepsilon_1$) and imaginary ($\varepsilon_2$) part of the dielectric function
obtained from fits to the measured Mg$_y$Ni$_{1-y}$ reflection between $0.062<\hbar\omega<3.5$ eV for some
selected compositions that are indicated in the legend. For comparison $\varepsilon_1$ and $\varepsilon_2$ of
pure metallic Ni are displayed from ref.~\onlinecite{Lynch}.}
\end{figure}
The free electron conductivity is described by the Drude term with $\tau$ the relaxation time, and the effect of
interband transitions by the $N$ Lorentz terms with $f_j$ the oscillator strength, $\omega_{0,j}$ the resonance
frequency and $\beta_j$ the broadening. Metallic Mg-Ni reflections can be modelled using the Drude term and only
one Lorentz oscillator that accounts for interband transitions. This gives relaxation times for the Mg-Ni system
typically in the order of $10^{-14}-10^{-15}$ s. The real and imaginary part of the dielectric function are
plotted in Fig.~\ref{MetalEps} for some selected compositions. For comparison, the dielectric function of pure
metallic Ni from ref.~\onlinecite{Lynch} is displayed. On increasing $y$, $\varepsilon_1$ first increases until
$y=0.4$ and then decreases towards $y=0.8$. $\varepsilon_2$ shows the opposite behavior, it first decreases until
$y=0.4$ and then increases towards $y=0.8$. As can be seen from Fig.~\ref{MetalEps}, on decreasing $y$ the
experimental $\varepsilon(\omega)$ well approaches the dielectric function of pure Ni.

\subsection{Electrical resistivity}\label{RhoMetallic} Optical properties showed indications for the formation of
the two intermetallic phases Mg$_2$Ni and MgNi$_2$ to be present in as-deposited Mg-Ni composition gradient
films. However, the arrangement of the atomic components on the crystal lattice, i.e., the degree of which the
phases are ordered remains to be specified. This structural property of the film appears to play a role on
hydrogenation and will be discussed in this section by means of the electrical resistivity.

\subsubsection*{DC Resistivity} DC resistivity measurements were carried out for Mg$_{y}$Ni$_{1-y}$ compositions
between $0\leqslant y\leqslant 1$ on glass substrates. Five partially overlapping composition gradient samples
were needed to cover this composition range. The experimental electrical resistivity of pure Mg ($y=1$) and pure
Ni ($y=0$) are $5.69\pm0.06$ and $10.59\pm0.09$ $\mu\Omega$cm, respectively, which is slightly higher than
literature values~\cite{HandbookChemPhys85}. This is probably due to a thin oxide layer on top of the film, to
impurities in the film and/or to structural imperfections due to the deposition technique.

\begin{figure}
\includegraphics[width=\linewidth]{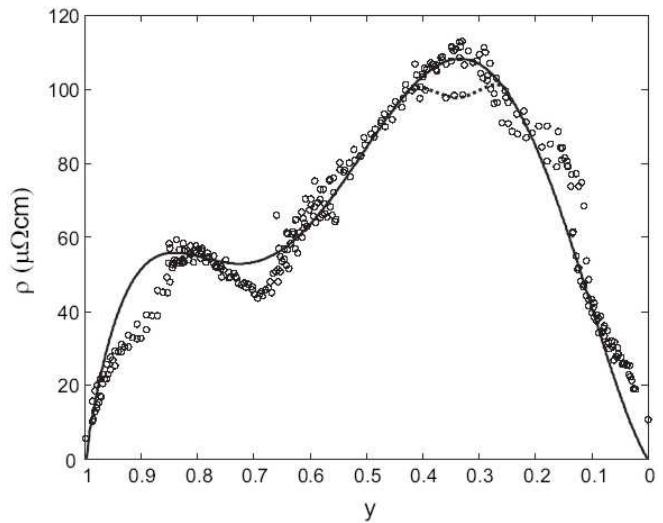}
\caption{\label{MetalResistivity} The composition dependence of the measured room temperature DC resistivity
$\rho_{\text{DC}}$ of Mg$_y$Ni$_{1-y}$. The solid line displays the fit based on Eq.~\ref{DCresistivity}. The
dotted line shows a guide to the eye for lowered resistivity values that are occasionally measured around
$y=0.33$.}
\end{figure}

Theoretically, the room temperature electrical resistivity of a concentrated binary metallic alloy system, which
consists of a single phase can be described by Nordheim's rule~\cite{Nordheim}. This predicts the resistivity to
be mainly caused by s-s electron transitions, implying the dependence on composition to be parabola-shaped with a
maximum around the equiatomic composition. This maximum resistivity indicates a disordered arrangement due to a
random mixture of the component atoms. However, solid solution systems involving transition metals like Ni,
generally do not obey Nordheim's rule. For these systems also s-d transitions contribute to the total
resistivity. In the transition metal poor region, the d band is filled which prevents transitions between the s
and d band and therefore the scattering probability is low, which points to low resistivity values. On increasing
the transition metal amount, the density of empty states into which an electron can be scattered increases and
consequently also the resistivity. The resistivity of such a single phase, totally disordered alloy can be
expressed as the sum of the partial resistivities due to s-s and s-d transitions, respectively:
\begin{equation}\label{DCresistivity}
    \rho_{\text{DC}}=A(1-y)y+B(p-y)^2(1-y)y^2,
\end{equation}
where $1-y$ is the atomic proportion of the transition metal and $A,B$ and $p$ are constants.~\cite{Mott}
Eq.~\ref{DCresistivity} describes a bell-shaped compositional dependence for which the maximum resistivity occurs
at a transition metal rich composition.

Fig.~\ref{MetalResistivity} shows the experimental room temperature DC resistivity of Mg$_y$Ni$_{1-y}$ together
with a fit based on Eq.~\ref{DCresistivity}, for which $A=0.248,B=6.51$ and $p=0.665$. For many compositions, the
fit curve nicely coincides with the highly reproducible experimental values, both showing the bell-shaped
composition dependence with an asymmetrically located maximum of about $108\ \mu\Omega$cm around $y=0.33$. As the
composition of the MgNi$_2$ phase is exactly centered around this maximum it indicates this phase to be
structurally disordered. Note further the significantly lowered resistivity compared to the fit between $y=0.6$
and 0.8. In several alloys of specific stoichiometric compositions, the unlike atoms can preferentially populate
distinct lattice sites creating an ordered arrangement of the atoms on the crystal lattice
sites.~\cite{Schroeder} The ordering processes result in a considerable reduction of the residual resistivity of
the alloy. As Eq.~\ref{DCresistivity} only assumes a single phase alloy, deviation of the fit with respect to the
experimental values is expected for these compositions. The resistivity dip is centered around the composition of
the Mg$_2$Ni phase, which means that contrary to MgNi$_2$, the Mg$_2$Ni phase has a well-ordered structure.

The maximum around $y=0.33$ has occasionally also been observed as a resistivity minimum, indicating the
formation of ordered MgNi$_2$ (see Fig.~\ref{MetalResistivity}). Typically, five identical glass substrates are
placed next to each other which are deposited simultaneously. In some cases, only one of the gradient samples had
the ordered MgNi$_2$ feature, whereas neighboring samples showed disordered MgNi$_2$. Whereas Mg-Ni gradient
samples that are deposited from the elements always show the ordered Mg$_2$Ni structure, the ordered MgNi$_2$
phase occurs only occasionally and seems to have a lower probability to be formed at room temperature. Since the
resistivity at $y=0.33$ can occur as a minimum or a maximum, the MgNi$_2$ composition is suspected to be an
order-disorder alloy. Based on systems like e.g. the Mn-Ni, which also possesses an ordered structure MnNi$_3$
comparable to Mg$_2$Ni in the Mg-Ni system, the Mg$_2$Ni composition is probably an order-disorder alloy as well.

Furthermore, large deviations with respect to the fitting curve are present around $y=0.15$ and 0.9, pointing
phases with a disordered and ordered arrangement, respectively.

\subsubsection*{Optical Resistivity}
\begin{figure}
\includegraphics[width=\linewidth]{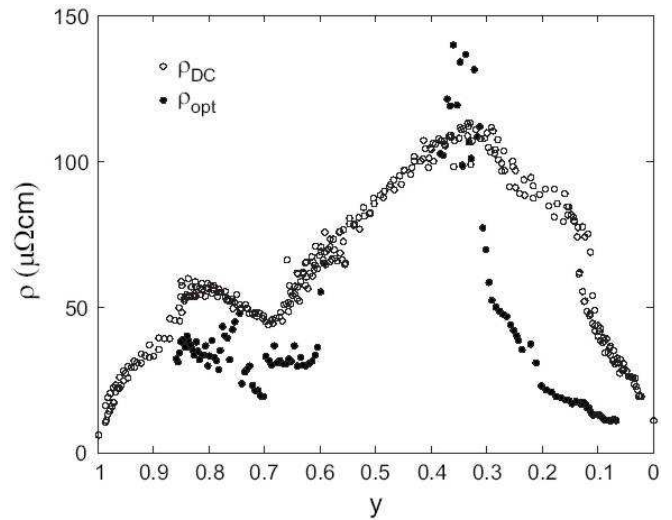}
\caption{\label{OpticalResistivity} The composition dependence of the optical resistivity $\rho_{\text{opt}}$ of
Mg$_y$Ni$_{1-y}$. Since the infrared part of the reflection spectra for $0.4\lesssim y\lesssim0.6$ could not
sufficiently well be fitted, the optical resistivity of this part is omitted. For comparison, the DC resistivity
$\rho_{\text{DC}}$ is depicted.}
\end{figure}
In order to check the mutual agreement between the electrical and optical properties, the DC resistivity
$\rho_{\text{DC}}$ can be compared to the optical resistivity $\rho_{\text{opt}}$ that is obtained from the Drude
fit parameters of the optical reflection spectra (see section \ref{MetalOptical}). $\rho_{\text{opt}}$ is related
to the plasma frequency and the relaxation time by
\begin{equation}\label{OpticRho}
    \rho_{\text{opt}}=\frac{1}{\varepsilon_0\omega_p^2\tau},
\end{equation}
where $\varepsilon_0$ is the vacuum permittivity, and is depicted in Fig.~\ref{OpticalResistivity}. Since the
infrared part of the reflection spectra for $0.4\lesssim y\lesssim0.6$ could not sufficiently well be fitted, the
optical resistivity in this range is omitted. Fig.~\ref{OpticalResistivity} shows the DC resistivity to be in
agreement with the optical resistivity. Around $y=0.33$ $\rho_{\text{opt}}$ coincides with $\rho_{\text{DC}}$
which indicates that also optical measurements show MgNi$_2$ to have a structurally disordered appearance on the
lattice. Moreover, it implies that the effect of disorder evaluated at $\omega=0$ and at optical frequencies
$\omega\sim10^{14}-10^{16}$ s$^{-1}$ is comparable. The optical resistivity further shows reduced values around
the composition of Mg$_2$Ni, though less pronounced than in the DC resistivity curve. Differences between
$\rho_{\text{opt}}$ and $\rho_{\text{DC}}$ may occur in this region where $\omega\tau\gtrsim1$, i.e., where one
period of the oscillating electrical field is of the same order of magnitude as the relaxation time. Both
electrical and optical properties are thus in good agreement with respect to the structurally disordered MgNi$_2$
phase.

\subsection{Structure}~\label{structure}
\begin{figure}
\includegraphics[width=\linewidth]{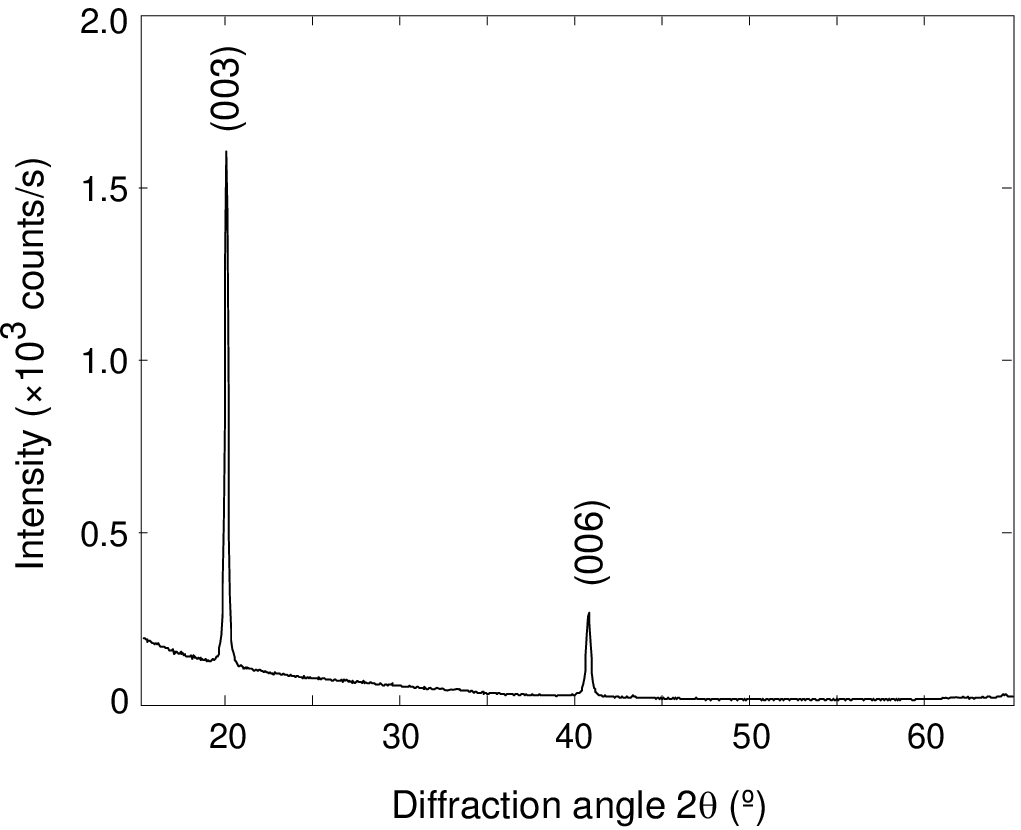}
\caption{\label{Mg2NiPresence} X-ray diffraction spectrum of as-deposited sputtered Mg$_y$Ni$_{1-y}$ on a Si
substrate for $y=0.68$. The Mg$_2$Ni (003) and (006) reflections show the c-axis of the hexagonal lattice to be
out of plane.}
\end{figure}
\begin{figure}
\includegraphics[width=\linewidth]{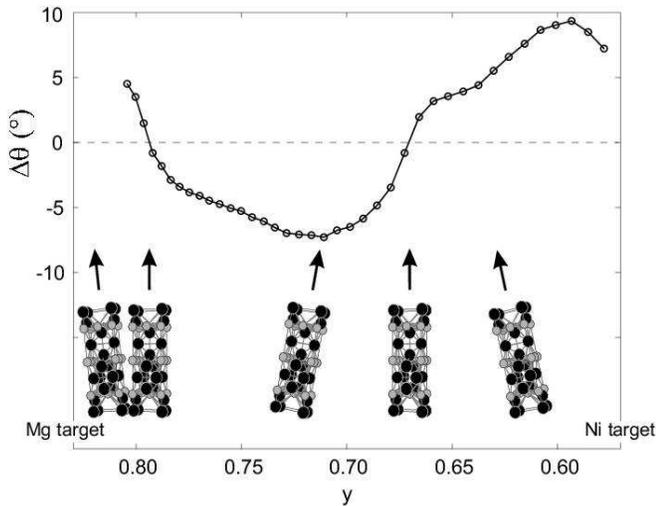}
\caption{\label{OrientationC-axis} Composition dependence of the orientation of the c-axis of the Mg$_2$Ni phase
in Mg$_y$Ni$_{1-y}$. $\Delta\theta$ is the deviation of the crystal axis with respect to the perpendicular to the
Si substrate, as illustrated for selected compositions by the cartoons of the hexagonal Mg$_2$Ni unit cell. The
location of both sputter targets is indicated.}
\end{figure}
In the previous sections, optical and electrical properties showed features of phase formation of Mg$_2$Ni and
MgNi$_2$ around their corresponding compositions. In this section, X-ray diffraction will establish the presence
of Mg$_2$Ni in Mg-Ni films that are sputter-deposited at room temperature. In the film layer, clusters of
crystalline Mg$_2$Ni form crystallites that are embedded in an amorphous metallic matrix of e.g. Mg$_2$Ni, Mg and
Ni. These structural properties of the film will be discussed for the region where Mg$_2$Ni can be detected.

X-ray diffraction measurements in the composition range $0.41<y<0.83$ indicate the presence of the Mg$_2$Ni phase
through its (003) and (006) reflections (see Fig.~\ref{Mg2NiPresence}), which are only observed between
$0.58<y<0.81$. Apart from Mg$_2$Ni, no other phases could be detected. Since the width of the (006) reflection is
approximately equal for all compositions, rocking curves of this reflection can be mutually compared. They show a
remarkable relation between $\theta$ at maximum intensity and the composition, as shown in
Fig.~\ref{OrientationC-axis}. From the deposition set-up one would expect the crystal axis at a given position on
the substrate to be directed to the most influential target seen from that position. On the Mg rich side, more Mg
is deposited than Ni so the Mg target would be presumed to dominate the deposition process and therefore also the
crystal axis direction. For the Mg richest compositions there is indeed a slight tilt $\Delta\theta$ towards the
Mg target (which defined as positive). At $y\simeq0.80$ the c-axis is perpendicular to the substrate and for
higher Ni concentrations it points to the Ni target, as would be expected. However, at $y\simeq0.68$ the c-axis
is again perpendicular to the substrate and for $y<0.68$ it even points towards the Mg target. If one assumes
that the particle beams have a homogeneous space distribution --- which is confirmed by the monotonic composition
gradient --- at the Ni rich side the Ni target would be expected to have a dominant effect on the crystal
orientation. Most probably the crystal axis is not only determined by the mutual particle beam currents but also
by particle migration on the film surface. Note, however, also the inexplicable correlation for $y=0.68$ and 0.80
between the fact that $\Delta\theta=0$ and the significantly enlarged transmission in the hydrogenated state (cf.
Fig.~\ref{IntroductionPicture}b).

\begin{figure}
\includegraphics[width=\linewidth]{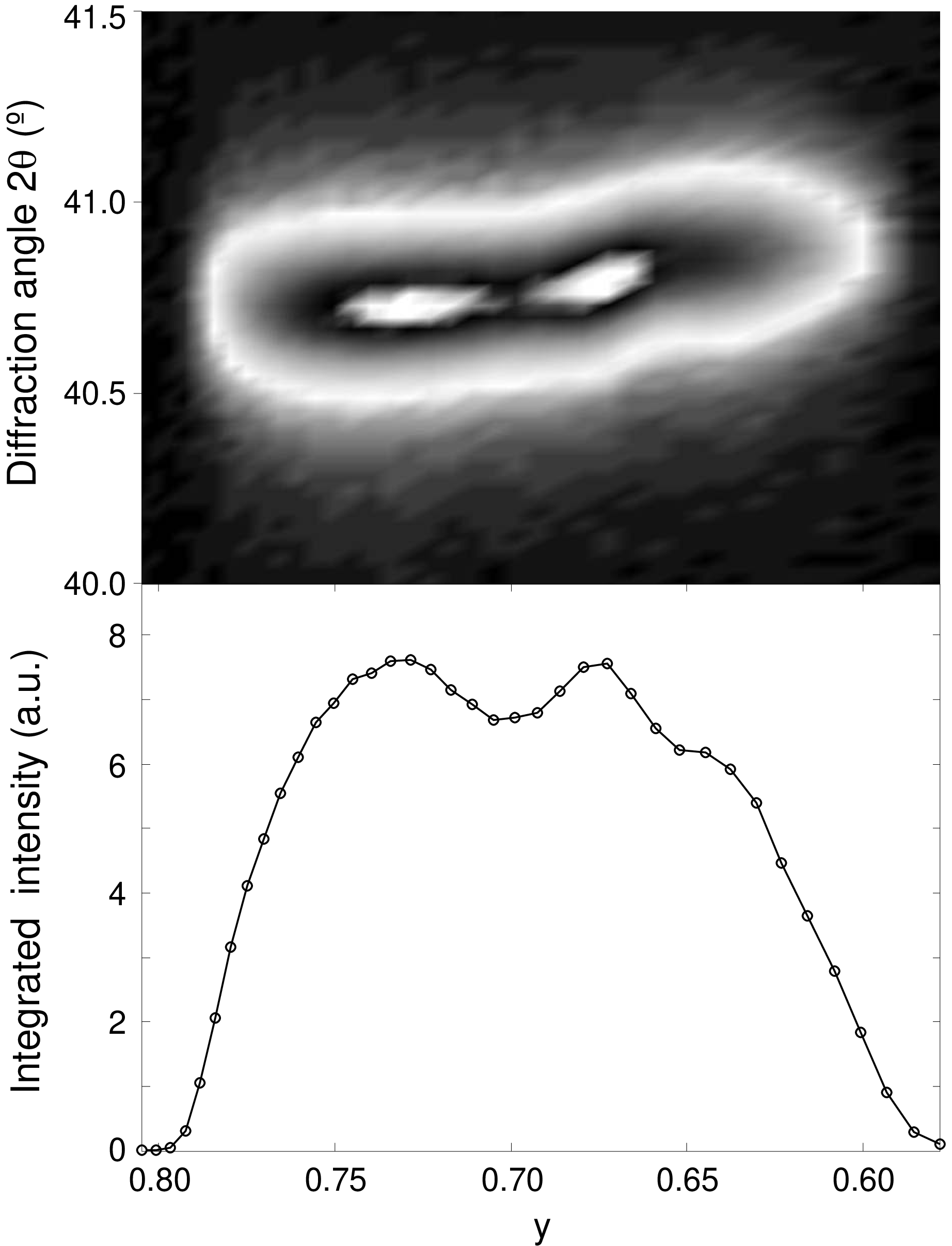}
\caption{\label{Mg2Ni006} Upper panel: the Mg$_2$Ni (006) reflection as a function of composition and diffraction
angle $2\theta$. $\theta$ is adjusted for the c-axis direction (cf. Fig.~\ref{OrientationC-axis}). Lower panel:
the composition dependence of the integrated (006) reflection, displaying the amount of crystalline Mg$_2$Ni in
Mg$_y$Ni$_{1-y}$.}
\end{figure}

Once having the direction of the Mg$_2$Ni crystal axis, the amount of Mg$_2$Ni can be determined as a function of
composition (Fig.~\ref{Mg2Ni006}). For this purpose, $\theta-2\theta$ scans are performed for the (003) and (006)
reflection with $\theta'=2\theta/2+\Delta\theta$, i.e., with $\theta$ adjusted for the tilt of the crystal c-axis
compared to the (400) reflection of the Si substrate. For angles $2\theta$ which showed a reflection peak due to
Mg$_2$Ni, the amorphous signal has been simulated based on regions with no distinguishable peak, which was then
subtracted from total reflection signal. Integration of the peak gives the amount of crystalline Mg$_2$Ni that
can be observed by XRD. As can be seen from Fig.~\ref{Mg2Ni006}, Mg$_2$Ni is present for $0.58<y<0.81$ though it
preferentially forms at Mg rich compositions. Although from the Mg-Ni phase diagram the highest amount of
Mg$_2$Ni is expected at one composition, $y=\frac{2}{3}$, Fig.~\ref{Mg2Ni006} shows two compositions with an
equal maximum amount Mg$_2$Ni, at $y=0.68$ and 0.73. It should also be noticed that only at $y=0.68$ the total
reflection signal is reduced by more than 30\%. This means that at this composition there is much less amorphous
Mg$_2$Ni than elsewhere, which makes it relatively the most crystalline composition.
\begin{figure}
\includegraphics[width=\linewidth]{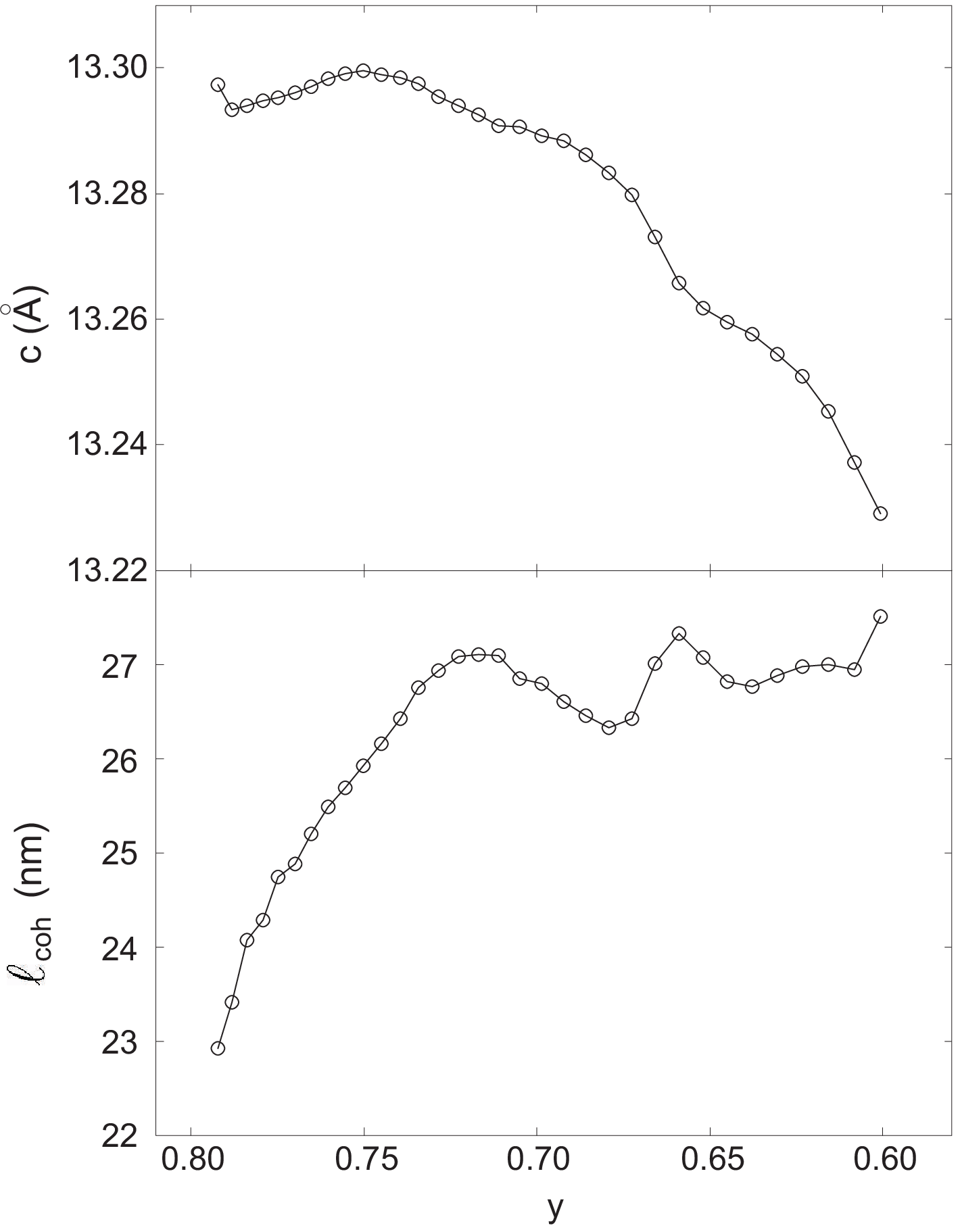}
\caption{\label{C-axisCohlength} Upper panel: the composition dependence of the c-axis lattice parameter $c$ of
Mg$_2$Ni in Mg$_y$Ni$_{1-y}$. Lower panel: the composition dependence of the coherence length $\ell_{\text{coh}}$
of Mg$_2$Ni clusters in Mg$_y$Ni$_{1-y}$.}
\end{figure}

The c-axis lattice parameter can be deduced from the peak position of the Bragg reflection at each composition
and is displayed in Fig.~\ref{C-axisCohlength}. The lattice mainly expands from $y=0.60$ to 0.76 after which it
gets smaller to Mg rich compositions. Since the molar volumes of Mg and Ni are 13.98 and 6.59 cm$^3$,
respectively~\cite{HandbookChemPhys85}, the expansion can be explained by the increasing amount of Mg next to
Mg$_2$Ni. At compositions where an elevated amount of Mg$_2$Ni is observed, $y=0.68$ and 0.73, the c-axis
distribution shows a slightly reduced lattice spacing, here also indicating the presence of a crystalline phase.

A measure of the size of Mg$_2$Ni grains in the c-direction is given by the coherence length $\ell_{\text{coh}}$,
which should be smaller or equal to the grain size. $\ell_{\text{coh}}$ can be estimated from the width of the
diffraction peak by the Scherrer equation~\cite{Scherrer},
\begin{equation}\label{Scherrer}
    \ell_{\text{coh}}=\frac{K\lambda}{\beta\cos\theta},
\end{equation}
where $\lambda$ is the emitted wavelength, $\beta$ is the peak width at half maximum (in radians), $\theta$ the
diffraction angle, and $K$ a factor which depends on the shape supposed for the intensity distribution of
diffraction. As a first attempt K is assumed to be 0.9. Fig.~\ref{C-axisCohlength} shows the coherence length in
the growth direction which increases from the Mg rich part to the Mg poor part from 23 to about 27 nm. Around
$y=0.68$ and 0.73 the coherence length is increased, indicating the formation of larger crystallites. In the film
layer, this implies on average an enhanced crystallinity due to Mg$_2$Ni which supports the observed elevated
amount of Mg$_2$Ni for both compositions.

\subsection{Surface morphology}
\begin{figure*}
\includegraphics[width=\linewidth]{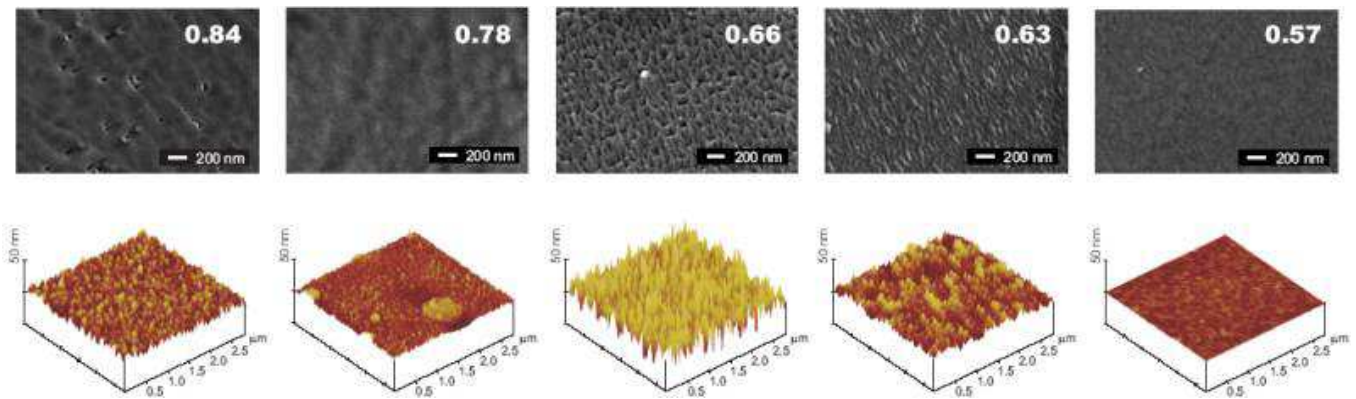}
\caption{\label{Surfaces} The surface morphology of Mg$_y$Ni$_{1-y}$ for some selected compositions that are
indicated in the right corners of the upper series images. For comparison surfaces recorded by SEM (upper series)
and AFM (lower series) are displayed together.}
\end{figure*}
X-ray diffraction revealed the structure of the as-deposited Mg-Ni thin films in the region that contained
crystalline Mg$_2$Ni. In order to complete the structural picture, in this section we will discuss the surface
morphology of as-deposited Mg-Ni films on Al$_2$O$_3$ substrates for compositions $0.55<y<0.85$.

\begin{figure}
\includegraphics[width=\linewidth]{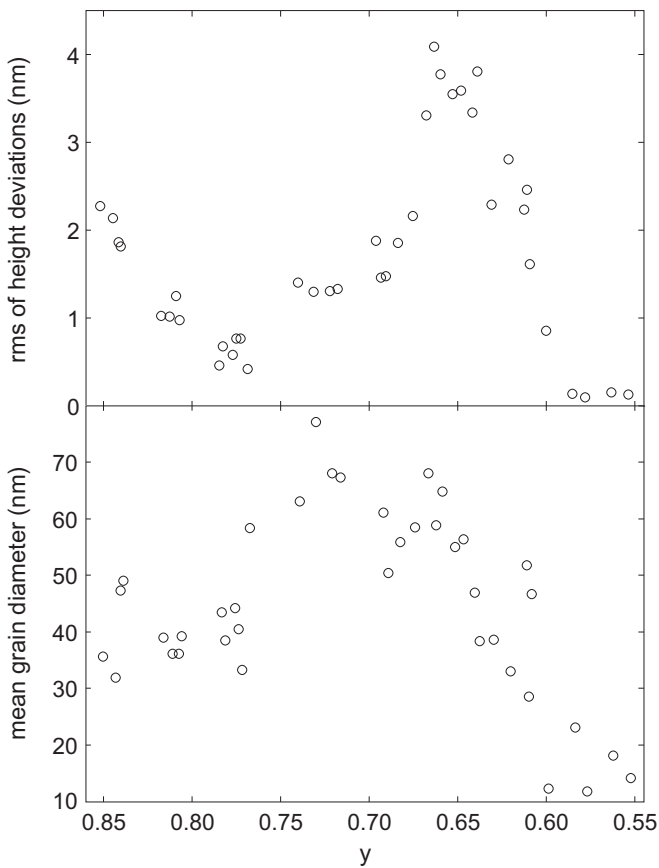}
\caption{\label{RoughnessM} Upper panel: the root-mean-square (rms) of height deviations at the surface of
Mg$_y$Ni$_{1-y}$ plotted versus composition. Lower panel: the composition dependence of the mean surface grain
diameter as determined from AFM measurements on Mg$_y$Ni$_{1-y}$ surfaces.}
\end{figure}

The surface morphology at some selected compositions is depicted in Fig.~\ref{Surfaces}. The SEM images show a
cauliflower-like pattern, except around 0.66 where the surface looks like a field of connected, sharp thorns,
suggesting an enhanced surface roughness. AFM, which has a much higher resolution, confirms this behavior. In
order to quantify the observed roughness, the root-mean-square (rms) of the height deviations is deduced from AFM
measurements and is plotted in Fig.~\ref{RoughnessM}. From $y=0.85$ to 0.78 the roughness linearly decreases from
2.3 nm to 0.6 nm, where the surface is rather flat. A maximum of 4 nm is attained around $y=0.65-0.67$, which is
in accordance with the SEM measurements. Below $y=0.60$ the roughness is less than 0.15 nm, which indicates an
almost completely flat surface.

Also from AFM measurements the surface grain size is determined, and is plotted in Fig.~\ref{RoughnessM}. Between
$y=0.78$ and 0.85 the mean grain diameter is $35-45$ nm, irrespective of composition. Between $y=0.65$ and 0.74
an elevated grain size is observed of about 60 nm, which decreases sharply for higher Ni contents.

In general, surface morphology can be intrinsic to the structure of the deposited composition, be due to
oxidation during deposition, be caused by post-growth surface oxidation or be due to the Pd cap layer. The latter
two reasons are possibilities as the films are capped with only 4 nm Pd that form islands instead of a closed
layer between which oxidation can occur. In order to trace the origin of the observed morphology, the oxygen
content was measured by RBS. At the substrate/film interface and in the film layer no oxygen has been observed
for the investigated range of compositions. At the surface, the oxygen content is completely independent of
composition and is on average $1.74\times10^{16}\pm0.92\times10^{16}$ atoms/cm$^2$. Also the Pd content has been
measured by RBS, which turned out to be composition independent too. One therefore can conclude that the observed
composition dependent surface morphology is completely inherent to the structure of the deposited composition.

The roughness and the grain size appear to have the same trend and show elevated values for $0.60\lesssim
y\lesssim0.75$. Since this range coincides with the composition region for which crystalline Mg$_2$Ni is observed
by XRD (see section~\ref{structure}), Mg$_2$Ni thus provokes rough surfaces. The kinetics of the hydrogenation
process is dependent on the surface roughness, which is therefore expected to be enhanced in the presence of
Mg$_2$Ni.

The structure of cross sections of the film has been investigated by SEM through looking at regions where the
film has been broken open, caused by externally applied stresses. Due to the deposition method a columnar
structure is observed in the as-deposited state. At the surface each column has a hemispheroidal top that
together form the cauliflower structure as seen from above. Therefore, the surface grain size, as discussed
earlier, can be considered to be the surface of these hemispheroidal tops.

\subsection{Discussion}\label{DiscussionMg-Ni}
\begin{figure}
\includegraphics[width=7.4cm]{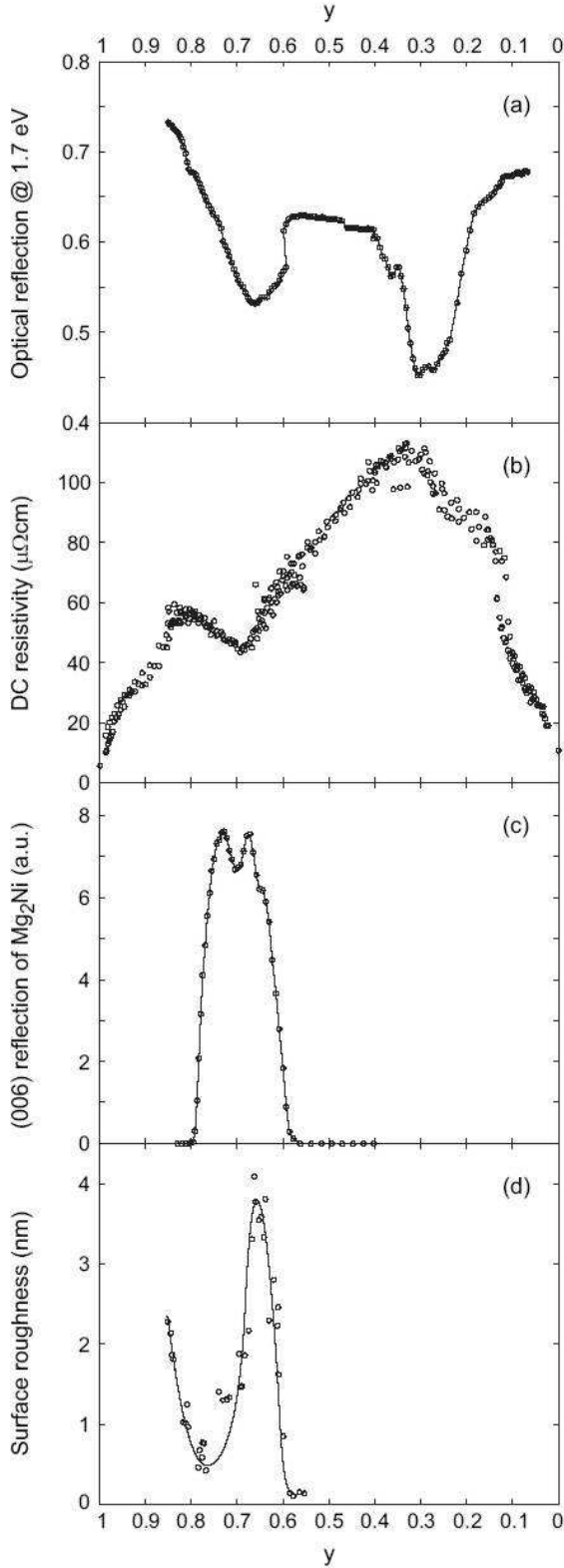}
\caption{\label{MetalDiscussion} Summary of the properties of the as-deposited metallic Mg-Ni system. The
presence of Mg$_2$Ni for $0.6<y<0.8$ is shown by: (a) the optical reflection, (b) the DC resistivity, (c) the
(006) Bragg reflection of Mg$_2$Ni and (d) the surface roughness displayed by the rms of height differences at
the surface. The presence of MgNi$_2$ around $y=0.33$ is indicated by the optical reflection and the DC
resistivity.}
\end{figure}
In the foregoing sections, as-deposited Mg-Ni composition gradient samples were subject to different techniques
in order to quantify the binary Mg-Ni alloy system. Fig.~\ref{MetalDiscussion} displays the optical, electrical,
structural and morphological properties as a graphical summary of the metal system. Optical reflection
measurements showed two minima that indicated the formation of the intermetallic alloys Mg$_2$Ni and MgNi$_2$
(Fig.~\ref{MetalDiscussion}a). At exactly the same composition ranges, the electrical DC resistivity showed a
site-ordered arrangement on the crystal lattice for the phase that was supposed to be Mg$_2$Ni and a structural
disordered arrangement for supposed MgNi$_2$ (Fig.~\ref{MetalDiscussion}b). The two regions of lowered optical
reflection can thus be explained by a site-ordered structure of Mg$_2$Ni which gives rise to a well-defined
interband absorption around $\hbar\omega=1.7$ eV, and by structurally disordered MgNi$_2$ for which various
interband absorption effects cause a strongly screened plasma frequency down to infrared energies. Further,
mutual agreement exists between optical and electrical measurements as they both indicate the presence of
Mg$_2$Ni and MgNi$_2$ and moreover show the latter one to be structurally disordered.

The presence of Mg$_2$Ni in a well-defined composition range was established by X-ray diffraction
(Fig.~\ref{MetalDiscussion}c). This range corresponds exactly to the composition region for which the optical
reflection and the DC resistivity showed lowered values, both already suggesting the Mg$_2$Ni phase. Remarkable
is the elevated presence of Mg$_2$Ni at two compositions, for $y=0.68$ and 0.73, instead of the single one at
$y=\frac{2}{3}$ that would be expected from the Mg-Ni phase diagram. Further, also the c-axis lattice size and
the coherence length of Mg$_2$Ni showed an enhanced crystal formation at both compositions.

The presence of Mg$_2$Ni causes the surface of the thin film layer to be much more rough than for neighboring
compositions, as can be concluded from SEM and AFM studies (Fig.~\ref{MetalDiscussion}d). Related is the surface
grain size which also shows elevated values if Mg$_2$Ni is present. Both morphological phenomena occur in a
composition range that exactly matches with the optical, electrical and structural anomalies of the Mg$_2$Ni
phase (cf. Fig.~\ref{MetalDiscussion}).
\begin{figure*}
\includegraphics[width=\linewidth]{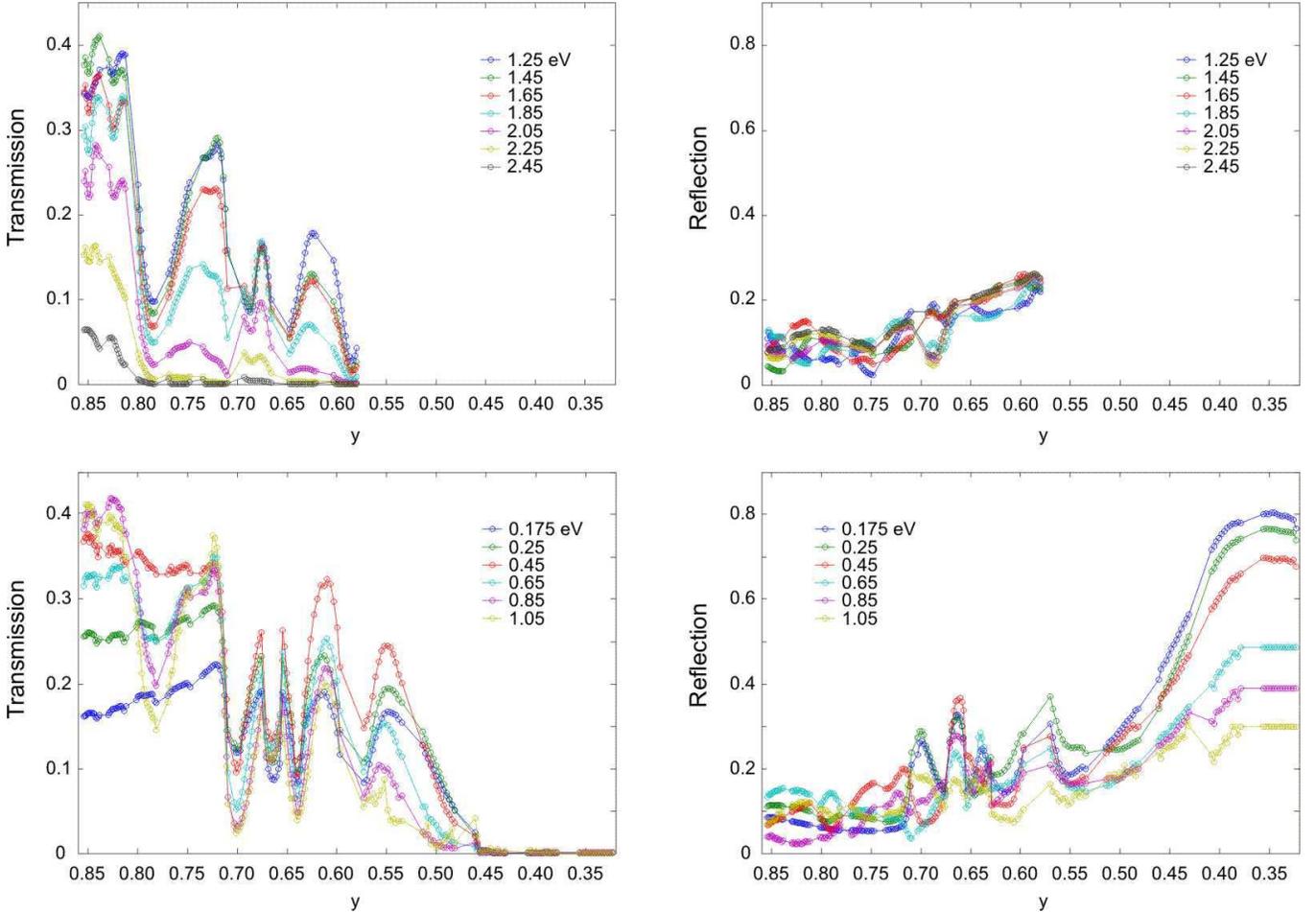}
\caption{\label{OpticalRT} The composition dependence of the optical transmission and reflection at visible and
infrared energies (indicated in the legend) of hydrogenated Mg$_y$Ni$_{1-y}$H$_x$ on CaF$_2$ substrates, covered
with a 4 nm thin Pd cap layer.}
\end{figure*}
Apart from these largely correlating aspects, there are also properties of the Mg-Ni composition gradient films
that are less well understood. For example, the DC resistivity shows a widely spread out hump around $y=0.15$ and
a wide dip around $y=0.9$ (see Fig.~\ref{MetalResistivity}) pointing to an increased structural disorder and
site-order, respectively. However, on the basis of the Mg-Ni phase diagram, no special features are expected for
both composition regions and no other measurements were performed at these compositions. Furthermore, the c-axis
orientation of the Mg$_2$Ni crystal (see Fig.~\ref{OrientationC-axis}) showed a very remarkable composition
dependence, which can not be explained by considering only the mutual power of the Mg and Ni target during the
deposition. Most probably also surface migration of the atoms should be regarded and possibly the direction of
the crystal axis is determined by the formed phases too.

\section{The M$\textrm{g}$-N$\textrm{i-H}$ system}~\label{MgNiH} The previously described metallic Mg-Ni system
will now be considered for the same composition gradient films after hydrogenation. The stripe-like pattern that
was observed in transmission by a CCD camera (Fig.~\ref{IntroductionPicture}b) is quantified by photometric
spectrometry. In order to understand the result, first a geometrical interference origin is considered.
Vibrational spectroscopy then demonstrates the presence of Mg$_2$NiH$_4$ and its distribution on composition.
Optical analysis further indicates the presence of MgH$_2$ at Mg rich compositions. For the Ni rich region, i.e.,
for $y<0.67$, electrical DC resistivity shows, in a framework of an effective medium theory (Bruggeman
approximation) and classical percolation theory, the influence of hydrogen around the equiatomic composition and
on the MgNi$_2$ phase. Eventually, X-ray diffraction measurements are performed to visualize the structure of the
hydrogenated Mg-Ni-H thin film.

\subsection{Optical properties}\label{OpticalPropertiesH}

The peculiar optical transmission of the hydrogenated Mg-Ni-H system was illustrated in the \textit{Introduction}
by means of a camera recorded image (Fig.~\ref{IntroductionPicture}b). In order to quantify the optical
properties, the transmission $T$ and reflection $R$ are measured by photometric spectrometry on composition
gradient films ($0.32<y<0.86$) on CaF$_2$ substrates. Fig.~\ref{OpticalRT} shows the very reproducible infrared
and visible transmission and reflection spectra for selected photon energies. The visible transmission again
displays the composition dependence that was earlier observed as a stripe-like pattern in the camera image.
However, for decreasing energies other narrow transmission maxima appear to develop around $y=0.65$ and 0.55.
Note the reduced intensity at $y=\frac{2}{3}$ corresponding to the composition of stoichiometric Mg$_2$NiH$_4$.
Remarkable is the absence of an enhanced transmission around $y=0.8$ where the camera image showed a bright
yellow stripe. As the spectrometer beam size is about 3 mm and since this region is only $0.5-1$ mm on the
sample, detection seems not to be possible. It should further be pointed out that the optical transmission
attains its maximum values at exactly $\hbar\omega=0.45$ eV for about the entire composition range. Moreover, at
the same energy, the composition region around $y=0.78$ changes from almost composition independent to strongly
dependent. Therefore, this energy might be a transition point for different absorbing processes, e.g. polaron
absorption below it and interband absorption above it, with minimum absorption effects at 0.45 eV. Eventually,
for Ni rich compositions the transmission falls off and is below 0.5\% for $y<0.45$.

The reflection generally shows the mirror image of the transmission, i.e., having a maximum at a transmission
minimum. This can be seen by the narrow composition regions which show an enhanced transmission as well as for
$y<0.45$ where the system approaches metallic properties and consequently the reflection increases sharply.
However, around $y=0.78$ the transmission shows a broad dip for $\hbar\omega>0.45$ eV which is not compensated by
an increased reflection, and consequently the absorption $A=1-R-T$ is high. This corresponds well to the dark
reddish region on the camera image.

It thus appears that Mg-Ni-H composition gradient thin films have a strong composition dependence. However, the
ternary Mg-Ni-H phase diagram~\cite{Zeng} only predicts the presence of MgH$_2$ at Mg rich compositions that
gradually changes to Mg$_2$NiH$_4$ at Mg poor regions, which can not account for the observed composition
dependence. In the following, analysis of the optical properties will lead to an understanding of the
transmission pattern.

\begin{figure}
\includegraphics[width=\linewidth]{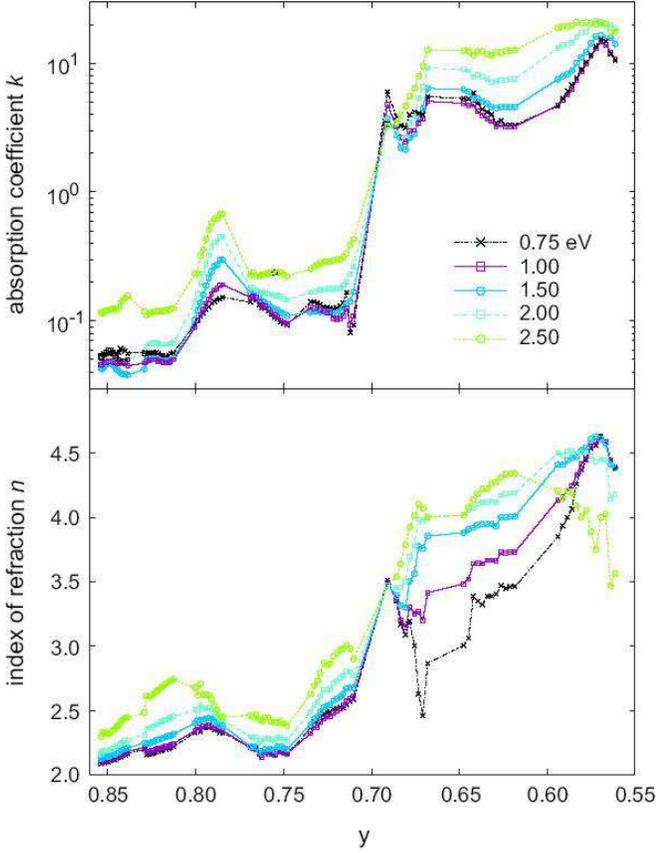}
\caption{\label{nk} The composition dependence of the absorption coefficient $k$ and the index of refraction $n$
of Mg$_y$Ni$_{1-y}$H$_x$ for some selected photon energies (indicated in the legend).}
\end{figure}

In general one would expect that a transparent wedge-shaped film were the thickness is of the order of the
probing wavelength $\lambda$, such as the Mg-Ni-H gradient samples, produces Fizeau fringes when studied in
reflection or transmission.~\cite{Hecht} Simulation of this interference effect with an overall index of
refraction of the Mg-Ni-H layer $n=3.5$ and an overall absorption coefficient $k=0.3$ ($n$ and $k$ based on
ref.~\onlinecite{VanMechelen} and this work, see Fig.~\ref{nk}), indicates the observed transmission pattern (cf.
Fig.~\ref{OpticalRT}) not to have an interference origin. In order to be caused by Fizeau interference, the
reflection and transmission maxima should move on varying the energy and therefore be separated by a distance
$d=\lambda/2\alpha n$, where $\alpha$ is the wedge angle, which increases up to 230 mm for $\lambda=7.1$ $\mu$m
(i.e., $\hbar\omega=0.175$ eV). Since all experimental reflection and transmission maxima are static on varying
the energy and separated by unequal distances that are fixed on changing $\lambda$, the composition dependence is
not due to interference but is intrinsic to the Mg$_y$Ni$_{1-y}$H$_x$ layer.

\begin{figure}
\includegraphics[width=\linewidth]{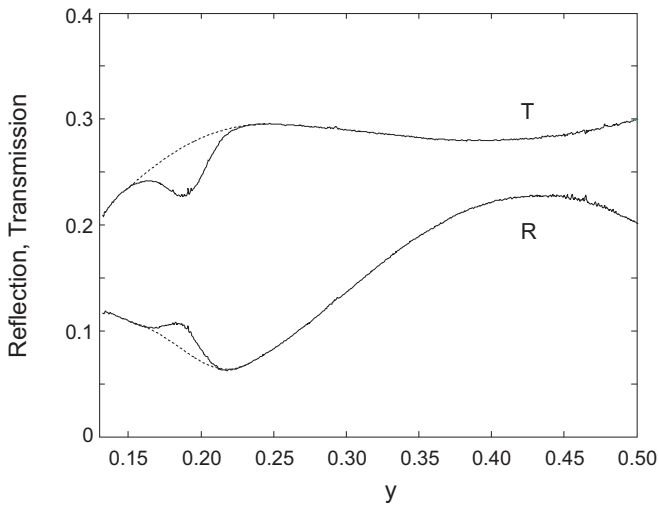}
\caption{\label{Phonons1} The optical transmission (T) and reflection (R) at $y=0.62$ in the hydrogenated state
(solid lines). The presence of the [NiH$_4]^{4-}$ phonon absorption centered around $\hbar\omega=0.197$ eV has
been made clear by plotting the simulated absorption-free signals (dotted lines) as well.}
\end{figure}

In order to identify the formed species in the Mg$_y$Ni$_{1-y}$H$_x$ layer, vibrational spectroscopy is performed
in the hydrogenated state. This revealed the presence of Mg$_2$NiH$_4$ by the infrared absorption modes of the
tetrahedral [NiH$_4]^{4-}$ cluster. Fig.~\ref{Phonons1} displays the transmission and reflection spectrum at
$y=0.62$ after hydrogenation. Simulated absorption-free signals are also plotted in order to point out the phonon
presence. The phonon absorption is centered around 1585 cm$^{-1}$ (i.e., 197 meV), which is close to literature
values~\cite{Richardson,Huang}, and has a substantial width of $250-300$ cm$^{-1}$ (i.e., $31-37$ meV),
indicating the Mg$_2$NiH$_4$ structure to be polycrystalline.

\begin{figure}
\includegraphics[width=\linewidth]{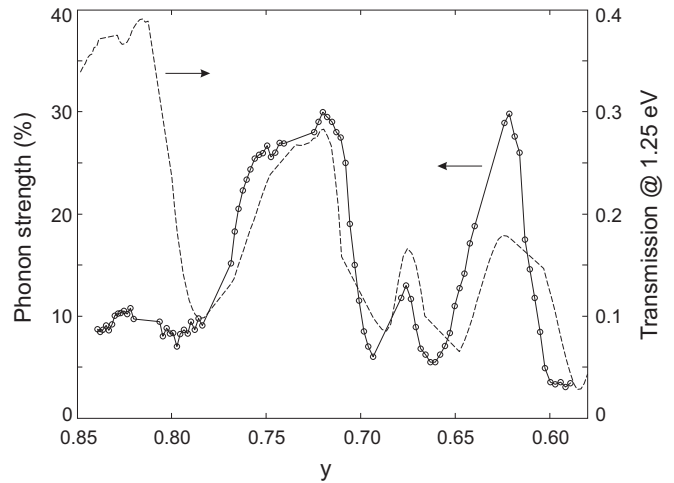}
\caption{\label{Phonons2} The composition dependence of the [NiH$_4]^{-4}$ phonon strength of hydrogenated
Mg$_y$Ni$_{1-y}$H$_x$. The strength is relative to the absorption-free signal level. For comparison, the optical
transmission at $\hbar\omega=1.25$ eV is shown.}
\end{figure}

Fig.~\ref{Phonons2} depicts the phonon absorption strength for compositions $0.59<y<0.84$ which is considered as
the height of the absorption compared to the absorption-free signal level. For this purpose, part of the measured
spectrum around $\hbar\omega=0.2$ eV where the absorption takes place has been removed, after which the entire
infrared spectrum is fitted in order to obtain the absorption-free signal around 0.2 eV. As can be seen from the
figure, for $y<0.78$ the phonon strength excellently coincides  with the optical transmission at
$\hbar\omega=1.25$ eV. This means that for $0.60<y<0.78$ the observed transmission is thus mainly due to
Mg$_2$NiH$_4$. On the Mg rich side, for $y>0.78$, the phonon strength is low whereas the transmission is rather
high. As Mg$_2$NiH$_4$ only slightly participates, in this region another transparent hydride should be formed.

Similar to the optical analysis for the binary metal alloy system (cf. section~\ref{MetalOptical}), the optical
constants of Mg$_y$Ni$_{1-y}$H$_x$ are deduced from the transmission and reflection measurements. A Drude-Lorentz
parametrization with 3 Lorentz oscillators is used to describe the complex dielectric function (cf.
Eq.~\ref{DrudeLorentz}). The reflection and transmission of the entire layered stack (CaF$_2$ --
Mg$_y$Ni$_{1-y}$H$_x$ -- PdH$_x$) are calculated using a transfer matrix method that considers the Fresnel
reflectance and transmittance coefficients at each interface and the absorption in each layer. The fitted
$\varepsilon$ fulfills the Kramers-Kronig relations and therefore the solution intrinsically satisfies causality.
After the fit procedure, comparison of the measured thickness of the metal layer with the thickness fit parameter
provides the expansion of the film during hydrogenation. This amounts to an expansion of $36\pm11$ \% in case of
fitting. Experimentally the expansion has been determined by profilometry on the same sample, which provided a
thickness increase of $27\pm4$ \%, without showing much composition dependence. Both expansions correspond well
to bulk which increases by 32 vol.\%~\cite{Schefer}.

From the fit parameters the index of refraction $n$ and the absorption coefficient $k$ of Mg$_y$Ni$_{1-y}$H$_x$
can be determined by the relationships $\varepsilon_1=n^2-k^2$ and $\varepsilon_2=2nk$, as plotted in
Fig.~\ref{nk}. For $y<0.8$ it was demonstrated that mainly Mg$_2$NiH$_4$ governs the system. In this region $n$
and $k$ have high values of $2.5-4.5$ and $0.2-20$, respectively, at $\hbar\omega=2$ eV. However, for
$y\gtrsim0.7$, the figure shows that $n$ and $k$ drop drastically down. The wide spread region of low optical
transmission around $y=0.78$ is characterized by elevated $k$ values which support earlier conclusions regarding
enhanced absorption. Extrapolation of $n$ and $k$ to $y=1$ indicates that they approach the literature values of
MgH$_2$ ($n=1.94-1.96$, $k=7.6\times10^{-3}$ at $2.107$ eV)~\cite{Isidorsson,Ellinger}. Whereas Mg$_2$NiH$_4$ is
mainly present for $y<0.8$ (cf. Fig~\ref{Phonons2}), MgH$_2$ thus dominates the optical properties for
compositions $y>0.8$. As MgH$_2$ is a very transparent semiconductor with an optical band gap of 5.6
eV,~\cite{Isidorsson} this fits well to the observed transmission (cf. Figs.~\ref{IntroductionPicture}b and
\ref{OpticalRT}).



\begin{figure}
\includegraphics[width=\linewidth]{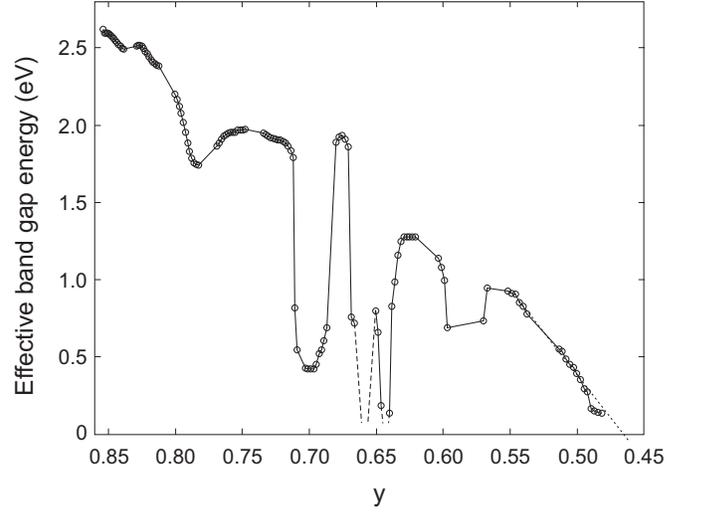}
\caption{\label{Bandgap} The composition dependence of the energy of the effective optical band gap $E_g$,
determined from the absorption coefficient $\alpha$ (see text). The dashed lines indicate the behavior of $E_g$
which decreases beyond the measured range. For $y<0.55$ the band gap energy is fitted by a linear relation
(dotted line) which crosses $E_g=0$ at $y=0.466$.}
\end{figure}

An effective optical band gap energy $E_g$ of Mg$_y$Ni$_{1-y}$H$_x$ can be determined from the energy dependent
absorption coefficient $\alpha$ that can be calculated from the transmission $T$ and reflection $R$ by the
Lambert-Beer law for an insulating thin layer:
\begin{equation}\label{LambertBeer}
    \alpha=-\frac{1}{t}\ln\left(\frac{T}{1-R}\right),
\end{equation}
with $t$ the thickness of the Mg$_y$Ni$_{1-y}$H$_x$ layer. The absorption coefficient is typically low below the
band gap energy and high above it. Therefore, $E_g$ is determined to be the energy at which $\alpha$ has its
maximum increase, which turns out to be between $5\times10^4$ and $6\times10^4$ cm$^{-1}$. Fig.~\ref{Bandgap}
shows also $E_g$ to have a strong composition dependence, which correlates well with the infrared transmission
for $\hbar\omega>0.5$ eV. At the Mg rich part, $E_g$ is large since MgH$_2$ dominates the system. The effective
optical band gap energy at $y=0.67$, i.e., the composition that is maximally determined by Mg$_2$NiH$_4$, turns
out to be 1.95 eV, which is slightly higher than literature values~\cite{Myers,Lupu}. At the Ni rich part, $E_g$
decreases almost linearly for decreasing $y$. Extrapolation of $E_g$ to $\hbar\omega=0$ gives $y=0.466$ at which
the Mg-Ni-H alloy has lost most of its insulating properties. For smaller $y$ the transmission is zero but up to
$y=0.35$ the reflection is still influenced by hydrogen, as can be seen from Fig.~\ref{OpticalRT}.

\subsection{Electrical resistivity}
\begin{figure}
\includegraphics[width=\linewidth]{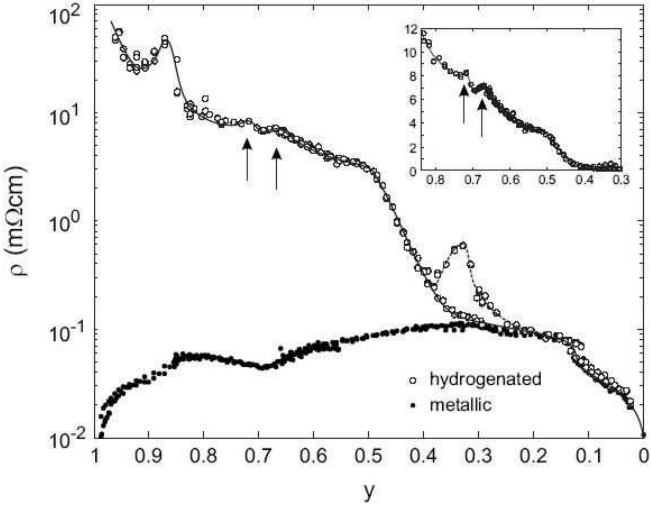}
\caption{\label{ResistivityH} Composition dependence of the DC resistivity $\rho$ of hydrogenated
Mg$_y$Ni$_{1-y}$H$_x$ compared to the resistivity of the as-deposited state. Elevated values around $y=0.67$ and
0.72 are indicated by arrows. Inset: linear plot of the composition dependence of the DC resistivity of
hydrogenated Mg$_y$Ni$_{1-y}$H$_x$ with emphasis on the maxima at $y=0.5,0.67$ and 0.72. The dotted line shows
the enlarged resistivity that is occasionally observed around $y=0.33$.}
\end{figure}
In the preceding section, optical analysis elucidated the optical transmission in the Mg rich region ($y>0.60$)
of the Mg-Ni-H system. However, the influence of hydrogen around the equiatomic composition and on the more Ni
rich side still remains to be evaluated. There is for instance a substantial infrared transmission around
$y=0.55$ though no Mg$_2$NiH$_4$ is present. This section considers the DC resistivity $\rho_{\text{DC}}$ of
hydrogenated gradient films for compositions $0\leqslant y<0.97$ on glass substrates. In order to demonstrate the
influence of hydrogen on compositions around $y=0.5$ and 0.33, the measured $\rho_{\text{DC}}$ is fitted to an
effective medium theory (Bruggeman approximation) and to classical percolation theory.

Fig.~\ref{ResistivityH} displays the DC resistivity of Mg$_y$Ni$_{1-y}$H$_x$ before and after hydrogenation. The
hydrogenated resistivity starts to deviate from the metallic resistivity for compositions $y\gtrsim0.2$. Below
this value, apparently the presence of hydrogen does not influence the lattice structure. For $y>0.2$ the
hydrogenated resistivity increases, and can be extrapolated up to about 100 m$\Omega$cm at $y=1$, which fits well
to the value of pure MgH$_2$. Around $y=0.68$ the metallic resistivity showed lowered values pointing to the
formation of site-ordered Mg$_2$Ni (section \ref{RhoMetallic}). After hydrogenation, two peaks are present around
$y=0.67$ and 0.72. In the semiconducting state, an enlarged resistivity indicates the formation of an ordered
arrangement on the crystal lattice. These two peaks therefore designate a site-ordered semiconducting phase
formation, which correlates with the  [NiH$_4]^{4-}$ phonon strength (cf. Fig.~\ref{Phonons2}) that also showed
maximum values for both compositions indicating an enhanced formation of Mg$_2$NiH$_4$. For the Mg rich region,
the DC resistivity behavior is thus in agreement with the optical properties.

Around the equiatomic composition the linear representation of the DC resistivity (see inset of
Fig.~\ref{ResistivityH}) shows a wide-spread hump pointing to increased resistivity values compared to
neighboring compositions. Around $y=0.33$ the hydrogenated system behaves in two ways: most frequently the
resistivity gradually drops for decreasing $y$, however, for about one third of the cases a large hump of 0.60
m$\Omega$cm is measured. Already in the as-deposited case this dual behavior was observed by showing a
resistivity maximum and sometimes a minimum (Fig.~\ref{MetalResistivity}). The lowered resistivity values in the
as-deposited state and the increased ones in the hydrogenated state, both point to the presence of a site-ordered
phase (see section~\ref{RhoMetallic}).

Supposed that between $y=0$ and 0.67 Mg$_2$NiH$_4$ is the only (or at least the dominant) insulator and Ni the
only metal. Compositions of them then form a composite system of semiconducting Mg$_2$NiH$_4$ and metallic Ni. In
order to see that the resistivity behavior around $y=0.5$ and 0.33 can not be described by such a system that in
composition changes between a semiconductor and a metal, we will apply two different models to the experimental
curves.

For a composite system of semiconducting and metallic clusters, the effective transport properties like the
resistivity can be calculated by Bruggeman's effective medium theory~\cite{Bruggeman,Landauer}. The effective DC
resistivity $\rho_{\text{DC}}$ of a mixture of Mg$_2$NiH$_4$ and Ni is than the solution of the self-consistent
relation,
\begin{equation}\label{Bruggeman}
    (1-x)\frac{\rho_{\text{DC}}-\rho_S}{\rho_{\text{DC}}+A\rho_S}+x\frac{\rho_{\text{DC}}-\rho_M}{\rho_{\text{DC}}+A\rho_M}=0,
\end{equation}
where $\rho_S$ and $\rho_M$ are the resistivities of semiconducting Mg$_2$NiH$_4$ and metallic Ni, respectively,
and $x$ is the metallic volume fraction. The geometric factor $A$ describes the shape of the metallic inclusions
in the composite system, and is related to the depolarization factor $D$ through $A=1/D-1$. For $D=1/3$ the
inclusions are spherical, whereas $D>1/3$ and $D<1/3$ correspond to prolate and oblate spheroids,
respectively.~\cite{Landau} The vertical axis of rotation of the inclusions are aligned parallel to the c-axis of
the metal lattice.

\begin{figure}
\includegraphics[width=\linewidth]{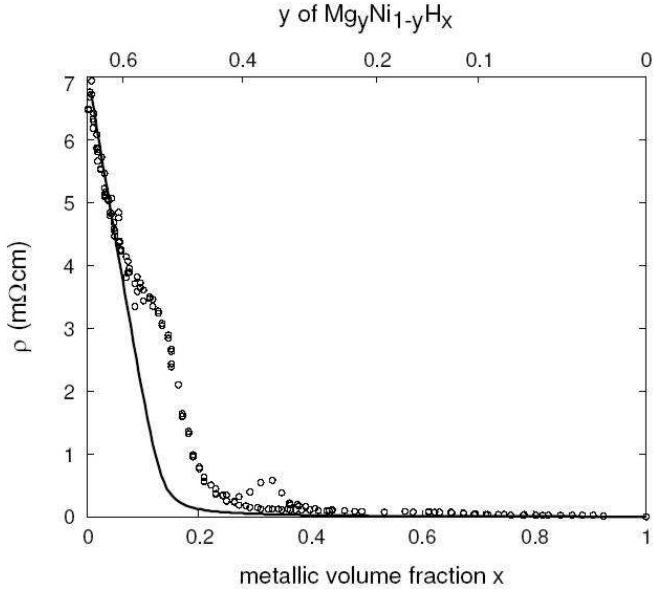}
\caption{\label{BruggemanFig} The electrical resistivity $\rho$ as a function of the metallic volume fraction Ni
$x$ in Mg$_2$NiH$_4$. The solid line displays the fit corresponding to Eq.~\ref{Bruggeman}.}
\end{figure}

Since Eq.~\ref{Bruggeman} is a monotonic function, it can not describe deviations like the humps we encountered
in the resistivity curve around $y=0.5$ and 0.33. However, recall that around $y=0.33$ the system behaves in two
ways (see Fig.~\ref{ResistivityH}), from which one represents a monotonic change of $\rho_{\text{DC}}$. For the
fitting procedure the latter one is included whereas the region around the equiatomic composition had to be
neglected. Fig.~\ref{BruggemanFig} shows the measured DC resistivity as a function of the metallic volume
fraction Ni together with the fit based on Eq.~\ref{Bruggeman} for a composite Mg$_2$NiH$_4$ -- Ni system. Here,
$\rho_S=7.11$ m$\Omega$cm, $\rho_M=10.59$ $\mu\Omega$cm, the mass density of Mg$_2$NiH$_4$~\cite{densityMg2NiH4}
is $2.706\times10^3$ kg/m$^3$ and that of Ni is $8.902\times10^3$ kg/m$^3$. The geometrical factor $A$ is easily
obtained as a fit parameter, $A=6.55\pm0.12$, which corresponds to a depolarization factor $D=0.132\pm0.002$. The
Ni inclusions in the composite mixture are thus oblate spheroids, as also found for the very related
Mg$_2$NiH$_{4-\delta}$ -- Mg$_2$NiH$_{0.3}$ composite system~\cite{Enache}.

The resistivity behavior around $y=0.5$ and $0.33$ (in case of the hump) can obviously not be described by
Bruggeman's effective medium theory which supposes a composite system that gradually changes between a
semiconductor and a metal. In order to support the view that in both composition regions different phase
formation occurs, classical percolation theory will also be applied to the measured resistivity.

The DC conductivity, $\sigma_{\text{DC}}=1/\rho_{\text{DC}}$, of a disordered metal-insulator mixture that
comprises a network of insulating and metallic clusters can be described by percolation.~\cite{Abeles} At the
percolation threshold $x_c$ a minimum metallic volume fraction generates a continuous conducting network
throughout the composite system which causes a steep increase in the conductivity.~\cite{Stauffer} Classical
percolation theory then describes the DC conductivity's dependence on the metallic volume fraction $x$ as
following a power-law behavior,
\begin{equation}\label{percolation}
    \sigma_{\text{DC}}\propto(x-x_c)^p.
\end{equation}

\begin{figure}
\includegraphics[width=\linewidth]{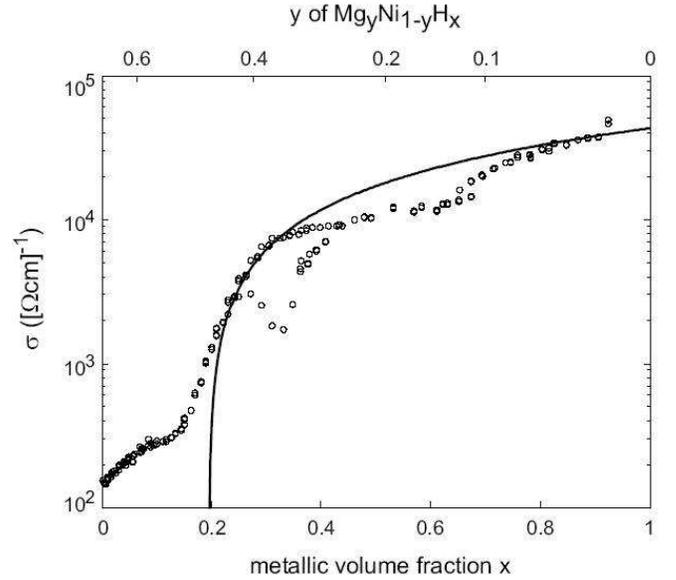}
\caption{\label{PercolationFig} The electrical conductivity $\sigma$ as a function of the metallic volume
fraction Ni $x$ in Mg$_2$NiH$_4$. The solid line displays the fit corresponding to Eq.~\ref{percolation}.}
\end{figure}

Similar to Bruggeman's effective medium model, Eq.~\ref{percolation} is a monotonic function of $x$ and can
therefore not describe local resistivity maxima (and thus conductivity minima) like the ones around $y=0.5$ and
0.33. Therefore, in the fitting procedure the monotonically changing resistivity around $y=0.33$ is included,
whereas the region around the equiatomic composition had to be omitted. Fig.~\ref{PercolationFig} shows the
measured electrical conductivity of the Mg$_2$NiH$_4$ -- Ni composite system versus the metallic volume fraction
and the fit based on Eq.~\ref{percolation}. It should be noticed that due to an irregularity of
$\sigma_{\text{DC}}$ around $x=0.66$ --- which is already present in the as-deposited state --- the fit does not
coincides with measured DC conductivity over the entire composition region. The percolation threshold $x_c$ is
$0.195\pm0.005$ (i.e., $0.443\pm0.005$ at.\% Mg) and the critical exponent $p$ is $0.96\pm0.02$. Site percolation
in an FCC lattice yields $x_c=0.198$ which is in excellent agreement with the obtained fit parameter. Further,
the percolation threshold at $y=0.443\pm0.005$ corresponds very well to the composition at which the effective
optical band gap $E_g$ equals zero, i.e., $y=0.466$ (cf. section~\ref{OpticalPropertiesH}), which both describe
the onset of metal behavior. The percolation critical exponent in 3D, $p$, is generally believed to be between 1
and 2.~\cite{Abeles,Stauffer,Kirkpatrick} The experimentally obtained fit parameter $p$ is very close to 1, which
is predicted by effective-medium theory.~\cite{Abeles}


Both Bruggeman's effective medium theory and classical percolation theory described well the overall DC
resistivity behavior apart from the two regions (around $y=0.33$ and 0.5) that showed an unexpected increase of
resistivity after hydrogen loading. This means that the behavior of both regions can not be described by a system
composed of a binary mixture of Mg$_2$NiH$_4$ and Ni. Nevertheless, hydride phase formation around $y=0.33$ is
suspected since similar to the hydride forming composition $y=0.67$, the resistivity in the as-deposited and
hydrogenated case is lowered and increased, respectively. Moreover, around $y=0.55$ a substantial infrared
transmission is present. For $y<0.60$, the [NiH$_4]^{4-}$ phonon strength is very low, which means that almost no
Mg$_2$NiH$_4$ is formed. Furthermore, the presence of MgH$_2$ especially at both compositions is very unlikely.
Concerning the equiatomic composition, in the metallic state, Orimo and Fujii~\cite{Orimo} observed amorphous
MgNi in bulk by X-ray diffraction, which on exposure to hydrogen becomes hydrogenated as demonstrated by
electrochemical measurements. Based on the just mentioned reasons, most likely Mg-Ni-H composition gradient
samples contain hydrogenated MgNiH$_x$ around $y=0.5$. Furthermore, optical measurements are performed down to
$y=0.32$, without showing any transmission for $y<0.4$. Therefore the increased resistivity around $y=0.33$ is
either due to a non-transparent hydride, most possibly MgNi$_2$H$_x$, or to a solid solution of hydrogen in
MgNi$_2$.

\subsection{Structure} Optical and electrical properties previously showed the distribution of Mg$_2$NiH$_4$, the
presence of MgH$_2$ and hydride formation around $y=0.5$ and $0.33$ in Mg-Ni-H composition gradient thin films.
This section considers X-ray diffraction on hydrogenated gradient samples on Si substrates in order to complete
the structural picture of the Mg-Ni-H system.

On exposure to hydrogen of an as-deposited Mg-Ni gradient thin film, the (003) and (006) reflection of Mg$_2$Ni
shift to lower diffraction angles, which indicates the formation of the solid solution phase Mg$_2$NiH$_{0.3}$.
If hydrogen uptake proceeds, the intensity of both reflections decreases to almost zero, as also seen for
Mg$_y$Ni$_{1-y}$H$_x$ thin films with $y=0.67$,~\cite{Lohstroh} and no other reflection appears. Even for
hydrogen loading at pressures up to $10^6$ Pa and temperatures of $100\ ^\circ$C no reflection of Mg$_2$NiH$_4$
and/or MgH$_2$ has been observed. Although, evidence for hydrogenation was obtained by the presence of optical
transmission. The absence of any reflection cannot be caused by the degree of crystallinity as vibrational
spectroscopy revealed strong phonon absorptions, but is rather due to the crystal size which might be below the
XRD detection limit. Nevertheless, high resolution line scans over the investigated compositions in the
hydrogenated state revealed two tiny reflections of the solid solution phase Mg$_2$NiH$_{0.3}$, that mainly exist
between $0.74<y<0.79$. The presence of Mg$_2$NiH$_{0.3}$ indicates poor hydrogenation for this composition range.

\begin{figure}
\includegraphics[width=\linewidth]{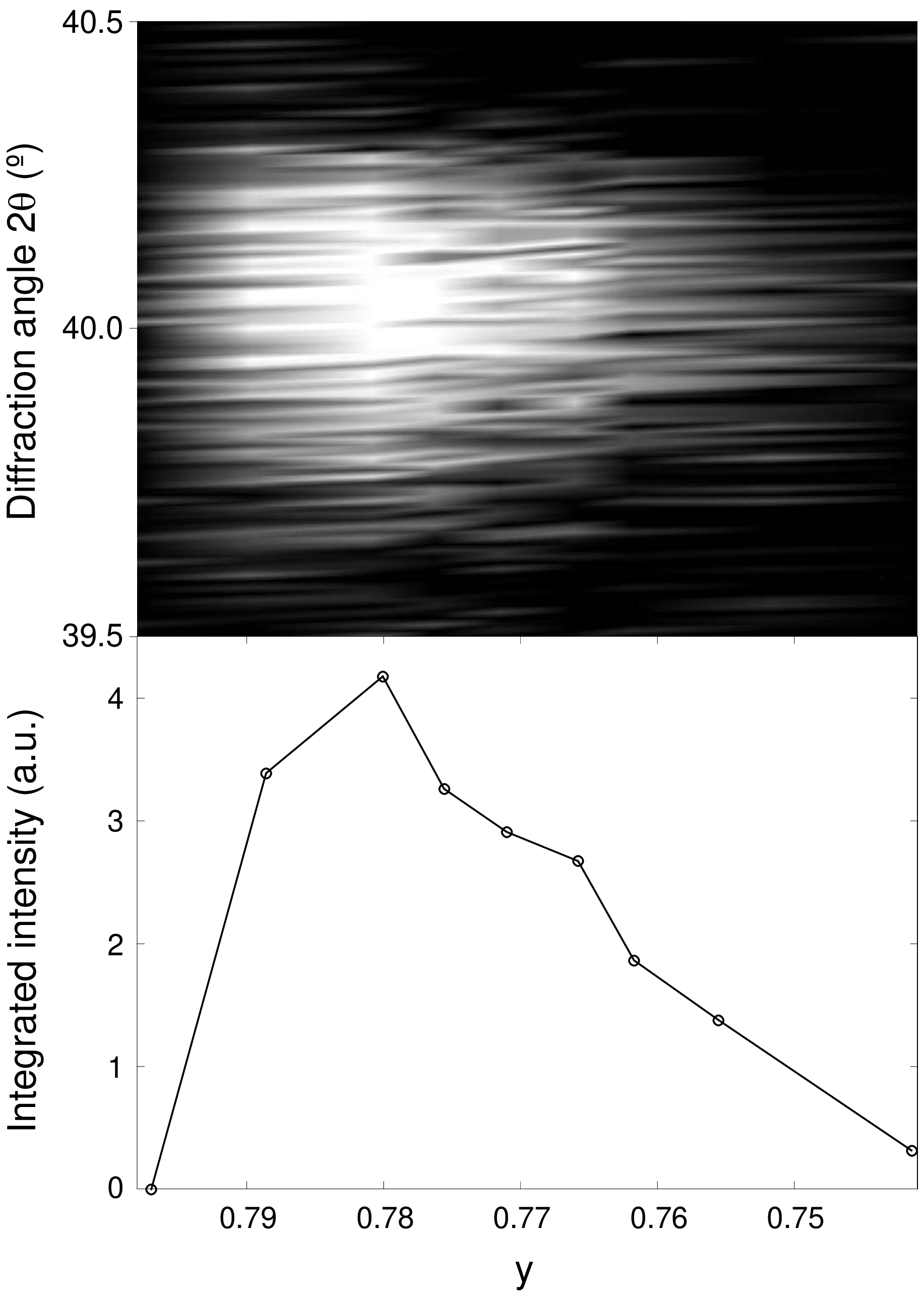}
\caption{\label{Mg2NiH0.3} Upper panel: the presence of the metal solid solution Mg$_2$NiH$_{0.3}$ (006)
reflection in hydrogenated Mg$_y$Ni$_{1-y}$H$_x$. Lower panel: the composition dependence of the integrated (006)
reflection displaying the amount of Mg$_2$NiH$_{0.3}$, which is maximum at $y=0.78$.}
\end{figure}

From $\theta-2\theta$ scans in which $\theta$ is adjusted for the orientation of the crystal c-axis, the amount
of Mg$_2$NiH$_{0.3}$ can be determined as a function of the deposited composition. The intensity distributions of
the (003) and (006) reflection show a maximum around $y=0.78$, as displayed in Fig.~\ref{Mg2NiH0.3}. Apparently,
$y\simeq0.78$ has the poorest hydrogenation of the entire investigated composition range. This agrees with
earlier conclusions based on the optical transmission and reflection and on the absorption coefficient $k$. As
the hydrogen uptake in the film is always hampered by a smooth surface, the AFM observed flatness in the
as-deposited state at this composition might be the reason of poor hydrogen loading.

\subsection{Discussion}\label{DiscussionMg-Ni-H}
\begin{figure}
\includegraphics[width=7.95cm]{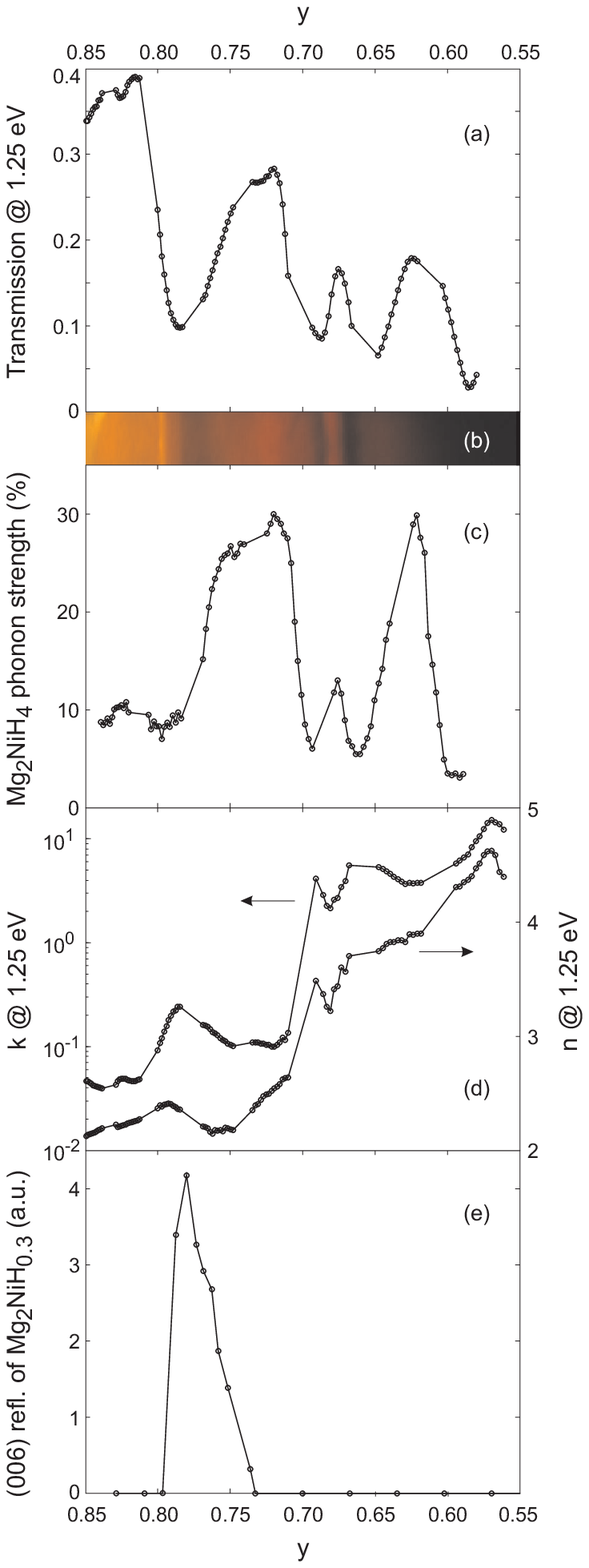}
\caption{\label{HydrideDiscussion} Summary of the properties of the Mg rich part of the hydrogenated Mg-Ni
system. (a) the optical transmission, (b) the optical transmission recorded by a CCD camera, (c) the strength of
the Mg$_2$NiH$_4$ phonon, (d) n and k at $\hbar\omega=1.25$ eV and (e) the integrated (006) reflection of
Mg$_2$NiH$_{0.3}$.}
\end{figure}

In the preceding sections, optical, electrical and structural properties of hydrogenated Mg-Ni-H composition
gradient thin films showed strong mutual correlations between different probes. This enabled the understanding of
the highly composition dependent transmission behavior which does not fit with the predicted Mg-Ni-H phase
diagram (see Fig.~\ref{IntroductionPicture}). Fig.~\ref{HydrideDiscussion} shows a graphical summary of the
physical properties of the Mg-Ni-H system. Transmission spectra recorded by photometric spectroscopy at visible
energies (Fig.~\ref{HydrideDiscussion}a) perfectly match with the transmission pattern of the camera recorded
image (Fig.~\ref{HydrideDiscussion}b). The elevated transmission at $y=0.80$, however, which has a bright yellow
appearance in the camera image, could be not probed due to its limited size on the sample. Extra optical contrast
is revealed by photometric spectroscopy in the infrared. Vibrational spectroscopy shows the distribution of
Mg$_2$NiH$_4$ as a function of composition (Fig.~\ref{HydrideDiscussion}c) that excellently coincides with the
optical transmission in the visible range. The observed transmission between $0.60<y<0.8$ is thus due to
Mg$_2$NiH$_4$.

For compositions $y>0.8$ the optical transmission and the optical band gap energy are large, whereas the
[NiH$_4]^{4-}$ phonon strength is low. However, in the same region, the index of refraction $n$ and the
absorption coefficient $k$, that are deduced from the optical transmission and reflection spectra, approach the
literature values of MgH$_2$ (Fig.~\ref{HydrideDiscussion}d). Since MgH$_2$ is a very transparent large band gap
semiconductor~\cite{Isidorsson}, it can be stated that above $y=0.8$ MgH$_2$ dominates the Mg-Ni-H system.

Eventually, around the equiatomic composition and at the more Ni rich region, i.e., for $y<0.60$, the electrical
DC resistivity showed elevated values around $y=0.5$ and 0.33 that could not fit within a composite system of
semiconducting Mg$_2$NiH$_4$ and metallic Ni. For $y=0.5$, this latter phenomenon together with lowered
resistivity values in the as-deposited state and a substantial infrared transmission in the hydrogenated state,
points to the formation of hydrogenated MgNiH$_x$. For $y=0.33$, however, no transmission has been observed.
Therefore, either a non-transparent hydride MgNi$_2$H$_x$ or a solid solution of hydrogen in the intermetallic
MgNi$_2$ phase is formed.

The stripe-like pattern that was shown by the camera recorded image can thus almost completely be understood from
the above mentioned properties. Although, the region around $y=0.78$, where the camera image showed a dark
reddish brown region does need some extra attention. The optical transmission and reflection spectra first showed
that the absorption $A=1-R-T$ in especially this region is high. The absorption coefficient $k$ displayed
elevated values pointing to the same conclusion. Eventually, X-ray diffraction measurements in the hydrogenated
state indicated an increased amount of the metallic solid solution Mg$_2$NiH$_{0.3}$ only around this composition
(Fig.~\ref{HydrideDiscussion}e). The high absorption is thus caused by poor hydrogenation, which gives rise to
the dark reddish brown appearance. It should be noticed that this region is exactly present in the composition
range where the system changes between MgH$_2$ and Mg$_2$NiH$_4$, i.e., around $y=0.78$. Most probably the poor
hydrogenation is due to the flat surface in the as-deposited state (cf. Fig.~\ref{RoughnessM}) which generally
slows down the hydrogen uptake.

The colors of the stripe-like pattern, shown in the CCD camera image, correspond well to the effective optical
band gap energy $E_g$. On increasing $y$ the hydrogenated film mainly changes from black to reddish towards
yellowish which is also displayed by $E_g$ that increases from 1 eV at $y=0.55$ (black visual appearance) to 2 eV
at $y=0.67$ (red) to 2.5 eV (yellow) around $y=0.85$. The narrow dark regions in between correspond to effective
band gap energies in the infrared range (see Fig.~\ref{Bandgap}). As also displayed by $E_g(y)$, the region
around $y=0.65$ has a very low effective band gap energy compared to neighboring compositions and therefore only
appears when probed with infrared energies. Remarkable is the [NiH$_4]^{4-}$ phonon strength which is low at this
composition. Therefore, further investigation is necessary to understand the origin of an enhanced transmission
at $y=0.65$.

In the \textit{Discussion} of the metallic Mg-Ni system (section~\ref{DiscussionMg-Ni}), we mentioned the
inexplicable lowered electrical resistivity around $y=0.9$. The DC resistivity in the hydrogenated state (see
Fig.~\ref{ResistivityH}) shows a huge maximum around the same composition. Similar to composition regions around
$y=0.33$ and 0.67 this points to the formation of a hydride phase. Unfortunately, no optical measurements were
performed in this composition range, and further research is necessary to unravel the origin of the behavior
around $y=0.9$.

\begin{figure}
\includegraphics[width=\linewidth]{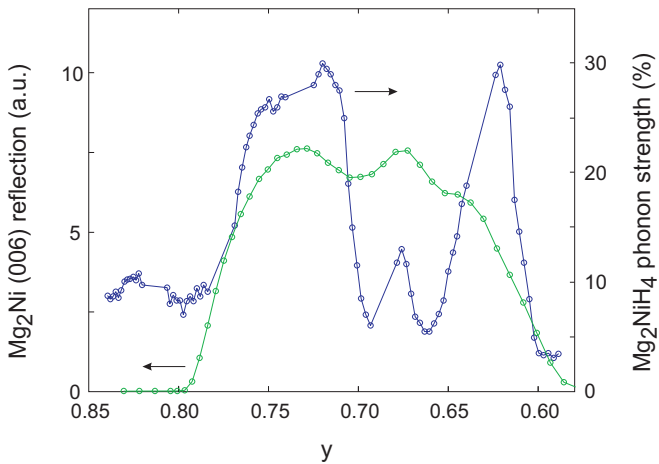}
\caption{\label{Metal-Hydride} The presence of Mg$_2$Ni shown by the total Mg$_2$Ni (006) reflection and the
Mg$_2$NiH$_4$ amount expressed by the Mg$_2$NiH$_4$ phonon strength. The resembling curves indicate the formation
of Mg$_2$NiH$_4$ preferentially from Mg$_2$Ni.}
\end{figure}

The relationship between metallic Mg$_y$Ni$_{1-y}$ and hydrogenated Mg$_y$NiH$_{1-y}$H$_x$ is given in
Fig.~\ref{Metal-Hydride} which represents the occurrence of Mg$_2$Ni and Mg$_2$NiH$_4$, respectively. The
Mg$_2$Ni (006) Bragg reflection shows that Mg$_2$Ni is mainly present between $0.6<y<0.8$. After hydrogenation,
the phonon absorption strength indicates that Mg$_2$NiH$_4$ is exactly present in the same composition range.
Moreover, the two compositions with a maximum amount of Mg$_2$Ni are after hydrogenation transformed into regions
that consist of an elevated amount of Mg$_2$NiH$_4$. This means that Mg$_2$NiH$_4$ preferentially forms from the
crystalline Mg$_2$Ni phase, instead of being formed from the elements.

\section{Conclusions} The behavior of the Mg-Ni-H system was studied by means of Mg$_y$Ni$_{1-y}$H$_x$ composition
gradient thin films. In the hydrogenated state, the optical transmission shows a stripe-like pattern of
alternating regions of low and high transmission pointing towards a high composition dependence. This phenomenon
is all the more unexpected as the predicted Mg-Ni-H phase diagram only represents a gradual change between two
transparent hydrides, Mg$_2$NiH$_4$ and MgH$_2$.

In order to understand the peculiar transmission pattern, first as-deposited metallic Mg-Ni thin films were
investigated. The optical reflection in the range $0.062<\hbar\omega<3.5$ eV indicated the formation of the
intermetallic phases Mg$_2$Ni and MgNi$_2$ in well-defined composition regions. The electrical DC resistivity
behavior showed both phases to have a site-ordered arrangement and structural disordered arrangement on the
crystal lattice, respectively. X-ray diffraction actually established the presence of Mg$_2$Ni and also showed a
maximum occurrence at two compositions instead of solely at $y=\frac{2}{3}$, as would be expected from the binary
Mg-Ni phase diagram.

In the hydrogenated state, vibrational spectroscopy showed the presence of Mg$_2$NiH$_4$ by the [NiH$_4]^{4-}$
absorption modes around $\hbar\omega=0.197$ eV. Its distribution as a function of composition indicates the
observed transmission between $0.60<y<0.8$ to be completely due to Mg$_2$NiH$_4$. For Mg rich compositions, i.e.,
$y>0.8$ optical analysis shows the system to have a strong MgH$_2$ character. Eventually, around the equiatomic
composition and for the more Ni rich region, i.e., $y<0.60$, the behavior of the electrical DC resistivity points
to hydride formation around $y=0.5$ and 0.33.

The stripe-like pattern that is observed in transmission at visible energies correlates well to many other
physical properties and turns out to be due to a very particular distribution of Mg$_2$NiH$_4$ together with the
presence of MgH$_2$ in the Mg rich region.

\section{Acknowledgements}
The work that is considered in this Master's thesis, has been performed in the Condensed Matter Physics group of
Prof. R.P. Griessen at the Vrije Universiteit Amsterdam, The Netherlands (2004). This experimental Master's
project has been performed with the help of many persons. First I would like to thank my father, L.F.M.P. van
Mechelen, who constructed the electrical resistivity apparatus. Second, but just as valuable, R. Gremaud who
assisted during a lot of measurements and with whom discussions helped the interpretation of the measurements.
Further, W.J. Lustenhouwer for the WDS composition measurements, S.M. Kars for the SEM measurements, H.
Schreuders and D.M. Borsa for the growth of the samples, J.H. Rector for performing RBS measurements, and N.P.
Armitage for proofreading the manuscript. Financial support from the Dutch Stichting voor Fundamenteel Onderzoek
der Materie (FOM) and the Stichting Technische Wetenschappen (STW) is acknowledged.


\begin{references}
\bibitem{densityMg2NiH4} The molecular volume is deduced from the experimental lattice parameter $a=6.49$ \AA\ of the Mg$_2$NiH$_4$ high
temperature phase from ref. \onlinecite{Schefer}.
\end{references}
\end{document}